\documentclass[fleqn]{2026SCGE}
\usepackage{bm}
\usepackage{algorithm}
\usepackage{amssymb}
\usepackage{amsmath}

\usepackage{physics}
\usepackage{mathtools}
\usepackage{graphicx}	                
\usepackage{bm}                         
\usepackage{enumitem}
\usepackage{url}
\usepackage{braket}
\usepackage{threeparttable}
\usepackage{placeins}

\newcommand\dex{\,{\rm dex}}
\newcommand\Kelvin{\,{\rm K}}
\newcommand\Mpc{\,{\rm Mpc}}

\newcommand\Kpc{\, {\rm {kpc}}}

\newcommand\kms{\,{\rm {km\, s^{-1}}}}
\newcommand\Msun{\,{\rm M_\odot}}

\newcommand\msunperyr{\, {\rm M_\odot}{\rm yr}^{-1}}

\newcommand\perccm{\, {\rm cm}^{-3}}

\newcommand\figem{\bf}                  
\newcommand\texteem{\bf\em}           
\def\softwarenamestyle[#1]{\textsc{#1}}

\begin{document}
\ensubject{subject}

\ArticleType{Article}
\SpecialTopic{SPECIAL TOPIC: }
\Year{2026}
\Month{January}
\Vol{69}
\No{1}
\DOI{??}
\ArtNo{000000}
\ReceiveDate{January 1, 1000}
\AcceptDate{April 6, 1000}

\title{Fountain pattern of baryon cycle revealed in galaxy ecosystems}{Fountain pattern of baryon cycle revealed in galaxy ecosystems}


\author[1,2]{Chengyu Ma}{}
\author[3,4]{Yangyao Chen}{{yangyaochen.astro@foxmail.com}}
\author[1,2]{Enci Wang}{{ecwang16@ustc.edu.cn}}
\author[5]{\\Houjun Mo}{}
\author[1,2]{Huiyuan Wang}{}
\author[3,4]{Tao Wang}{}
\author[6]{Qirong Yuan}{}
\author[7]{Min Du}{}
\author[3,4]{Zhaozhou Li}{}
\author[8,9]{\\Kai Wang}{}
\author[1,2]{Cheqiu Lyu}{}
\author[1,2]{Haoran Yu}{}
\author[1,2]{Zeyu Chen}{}
\author[7]{Xiaoxuan Chen}{}

\AuthorMark{Ma C}

\AuthorCitation{Ma C, Chen Y, Wang E, et al}


\address[1]{Department of Astronomy, University of Science and Technology of China, Hefei, 230026, China;}
\address[2]{School of Astronomy and Space Science, University of Science and Technology of China, Hefei, 230026, China;}
\address[3]{School of Astronomy and Space Science, Nanjing University, Nanjing, 210093, China;}
\address[4]{Key Laboratory of Modern Astronomy and Astrophysics, Nanjing University, Ministry of Education, Nanjing, 210093, China;}
\address[5]{Department of Astronomy, University of Massachusetts, Amherst, 01003, USA;}
\address[6]{School of Physics and Technology, Nanjing Normal University, Nanjing, 210023, China;}
\address[7]{Department of Astronomy, Xiamen University, Xiamen, 361005, China;}
\address[8]{Institute for Computational Cosmology, Department of Physics, Durham University, Durham, DH1 3LE, UK;}
\address[9]{Centre for Extragalactic Astronomy, Department of Physics, Durham University, Durham DH1 3LE, UK}


\abstract{
Baryons in galaxy ecosystems are believed to undergo continuous cycles of inflow 
and outflow, forming fountain-like patterns that encode key information about how 
galaxies acquire matter from their environments and respond through feedback. The 
presence of such baryon cycles has been inferred from pieces of observational 
evidence, but a concrete understanding remains elusive because individual galaxy 
ecosystems are diverse and dynamic. Here we introduce a stacking method that combines 
baryonic fields across ensembles of individual galaxy ecosystems to suppress irregularities 
and reveal the underlying pattern. 
Applied to a cosmological hydrodynamic simulation, this approach unveils 
strikingly regular patterns in gas properties across the full spatial extent 
of galaxy ecosystems, in close agreement with those inferred from 
observations. This method is straightforward to implement, allowing the processes shaping the 
gas-cycling pattern to be fully understood within the structure-formation paradigm, 
and a solid framework to be constructed for linking simulated galaxy ecosystems 
with observations.}


\maketitle


\begin{multicols}{2}

\section{Introduction}
\label{sec:intro}

In the current lambda-cold dark matter ($\Lambda$CDM) paradigm,
galaxy formation is an interplay between the two sectors
of the Universe \cite{peeblesLargescaleStructureUniverse1980,moGalaxyFormationEvolution2010}: the dark sector, composed of collisionless dark matter, and the luminous sector, composed of baryonic matter. 
A key characteristic of baryons around galaxies is that they must undergo
continuous cycles of inflow and outflow across multiple scales, forming a 
fountain-like pattern \cite{shapiroConsequencesNewHot1976,tumlinsonCircumgalacticMedium2017,veilleuxGalacticWinds2005,
wrightBaryonCycleModern2024,foxGasAccretionGalaxies2017,
zhangInspiralingStreamsEnriched2023} that encodes the 
operation of relevant physical processes.
Such a pattern, together with the context provided by the dark matter, 
forms a fundamental picture now known as the ``galaxy ecosystems'' 
(see ref.~\cite{tumlinsonCircumgalacticMedium2017} for a review).

Observational evidence has been accumulated to support the existence of
the baryon cycle, to quantify its properties, and to infer its driving 
mechanisms \cite{kimmAreColdFlows2011,
fabianObservationalEvidenceActive2012,
rupkeMultiphaseStructurePower2013,zhangInspiralingStreamsEnriched2023}. 
Along the same line, significant efforts have been made to build theoretical models
\cite{keresHowGalaxiesGet2005,dekelColdStreamsEarly2009,
pillepichSimulatingGalaxyFormation2018,nelsonFirstResultsTNG502019,
moTwophaseModelGalaxy2024,chenTwophaseModelGalaxy2025a} 
that have linked the discrete pieces of observational evidence 
into a self-consistent picture within the $\Lambda$CDM paradigm.
Despite the great advances in observational and theoretical studies, some 
fundamental questions remain to be answered:
the basic pattern of baryon cycle (e.g. Figure~1 of ref.~\cite{tumlinsonCircumgalacticMedium2017}),
remains conceptual, rather than being quantitatively derived from observations 
or simulations; it has not yet been clarified whether such a pattern
exists in or around all galaxies throughout their histories, or just 
reflects a time-averaged or ensemble-averaged behavior;
the recycled 
inflow has been inferred to be necessary to accommodate the indirect 
constraints from observations, but to what spatial extent, in what angular configuration, and at what rate such recycling proceeds remain unclear;
the regimes of galaxy ecosystems, such as the interstellar medium (ISM), 
the circumgalactic medium (CGM), and the intergalactic medium (IGM),
are widely adopted in the literature to conceptually separate the scales of 
the baryon distribution, but definitions of these regimes have been 
too ambiguous, and often vary among specific projects, so that 
a comparison between results from different works is difficult.
These unsolved questions reflect the lack of a physically motivated, yet precise
picture that encapsulates the pattern of baryon cycle, on top of which 
clear definitions of the key concepts can be established and be adopted
consistently by the community. 

A major challenge to concretize the picture of baryon cycle
is the diversity of galaxy ecosystems and irregularities in individuals: 
galaxy populations can differ significantly in their properties;
for an individual galaxy,
the baryon distribution often exhibits a disordered, time-varying 
pattern, deviating from a stable, featureless disk or spheroid 
configuration \cite{pillepichSimulatingGalaxyFormation2018,nelsonFirstResultsTNG502019}.
These indicate that the conceptualized pattern of baryon cycle 
may not be precisely applicable to individual systems, but rather 
a statistical behavior of galaxy ensembles.

In this paper, we propose to reveal the precise pattern of baryon cycle
in a statistical manner by stacking the baryon fields
around ensembles of galaxies. 
As expected from statistical 
principles \cite{bishopPatternRecognitionMachine2006}, if an ensemble is governed  
by a coherent pattern and if the ensemble is sufficiently representative, 
stacking can suppress the irregularities of individuals to reveal the underlying 
pattern.
Such a stacking method has been adopted in observations to extract certain 
aspects of baryon cycle from noisy data \cite{limGasContentsGalaxy2018,popessoXrayInvisibleUniverse2024,
zhangHotCircumgalacticMedium2024}, but its 
power was overlooked in hydrodynamical simulations.
As we will show by applying the method to a cosmological hydrodynamical
simulation, the statistical pattern of baryon cycle exhibits striking regularities 
that allow us to unambiguously define the regimes of galaxy ecosystems, 
precisely quantify the baryon properties within each regime, and understand the 
physical processes shaping the pattern.

This paper is organized as follows.
In \S\ref{sec:method}, we introduce the stacking method and 
the simulation used to demonstrate the method.
In \S\ref{sec:results}, we analyze the pattern of baryon cycle revealed 
by the stacking.
In \S\ref{sec:summary}, we summarize our findings and discuss 
future directions.
Throughout this paper, we adopt a flat $\Lambda$CDM cosmology with parameters 
derived from the Planck 2015 results \cite{planckcollaborationPlanck2015Results2016}.

\begin{figure*}[t]
	\centering
	\includegraphics[width=\textwidth]{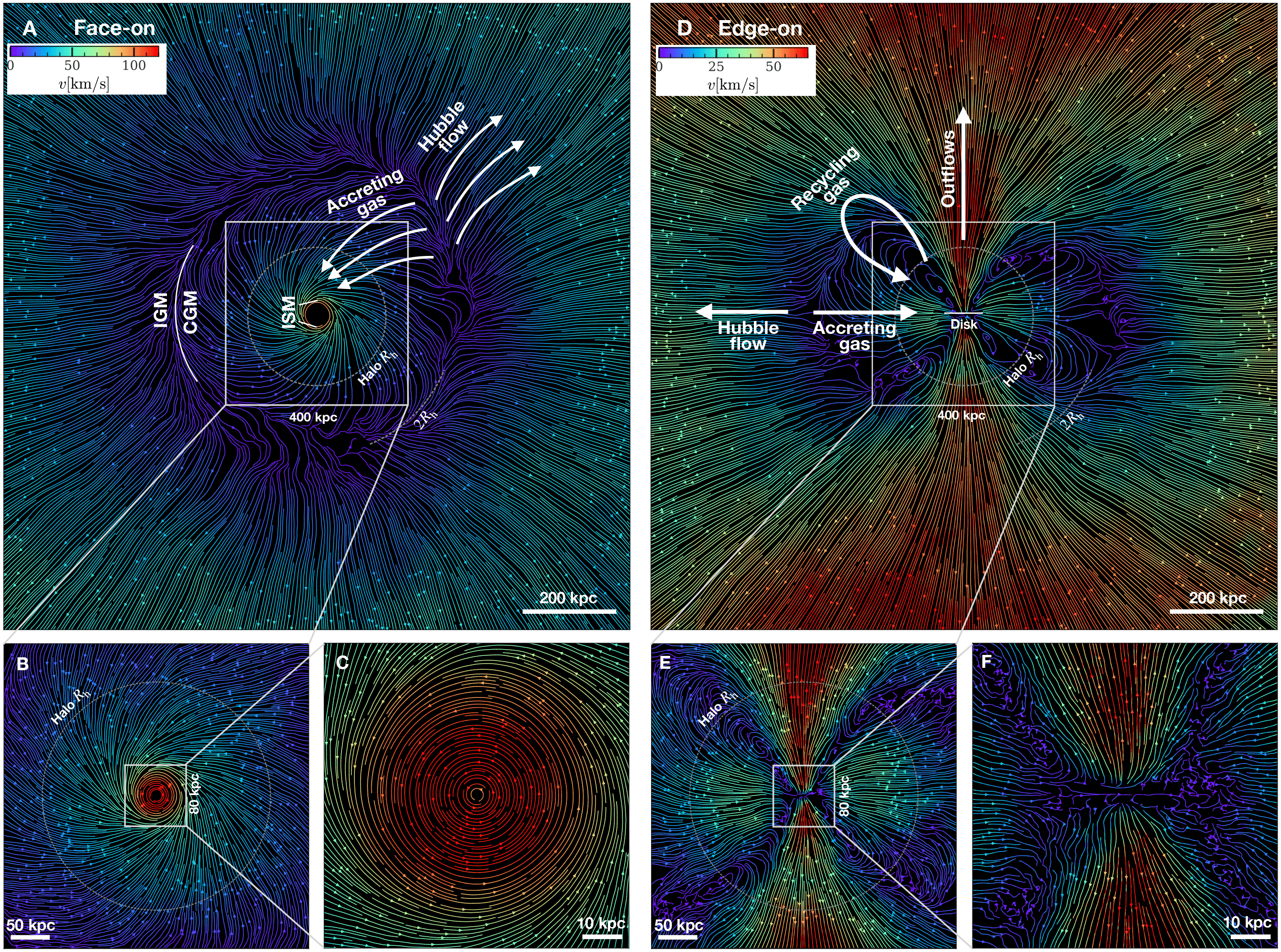}
	\caption{
		{\figem Stacked velocity field of gas around an ensemble of galaxies.} 
		The field was obtained for the fiducial ensemble (star-forming galaxies 
		with $M_* \approx 10^{10} \Msun$ at $z=0$). 
		{\figem (A)}--{\figem (C)} 
		Face-on view for a slice along the disk plane.
		{\figem (D)}--{\figem (F)}
        Edge-on view for a slice perpendicular 
		to the disk plane.
		Each slice has a thickness of $20\Kpc$ centered on the centers of the 
		galaxies in the ensemble.
		Velocity is projected onto the slice plane.
		Streamlines show the velocity field,
		with arrows indicating the directions and colors indicating the 
		magnitudes.
		Three zoom-in levels for each slice are shown.
		Dashed circles mark the median virial radius ($R_{\rm h}$) of host halos.
		A smooth, coherent pattern is unveiled by the stacking,
		supporting the use of hydrodynamic simulations with realistic initial 
		conditions to recover the pattern governing ensembles 
		of galaxy ecosystems.
		}
		\label{fig:streamline}
\end{figure*}

\section{Method and data}
\label{sec:method}

\subsection{The stacking method}
\label{ssec:stacking-method}

Our stacking method takes an ensemble of galaxies as input, and constructs a composite field of a target property.
To maximize the generalizability,
we divide our stacking method into three abstract steps: 
(i) {\texteem Alignment}: align the fields of individual galaxies to maximize 
their similarity.
(ii) {\texteem Discretization}: aggregate resolution elements 
(e.g. particles or cells) from a galaxy 
in a hydrodynamic simulation into ``super-cells'' at specified locations, 
with each super-cell 
characterized by properties averaged over its constituent elements. 
This discretization yields a representation of the field that is 
independent of the resolution scheme (see \S1 of ref.~\cite{hopkinsNewClassAccurate2015} for a review). 
(iii) {\texteem Reduction}: apply a reduction operation on the super-cells
across the aligned ensemble to construct a composite field.

Formally, to construct the composite field ($f$) in 3D configuration space 
for a target property (scalar, vector component, etc.)
across an ensemble ($E$) of galaxies, the alignment step is performed by
establishing a local coordinate frame centered on each galaxy $g \in E$, 
so that the distribution of the target property measured in this frame
exhibits maximal similarity across galaxies in the ensemble.
The choice of the local frame depends on the sample being analyzed
and the target property being studied, which we will specify in 
\S\ref{ssec:sample}.
Around each galaxy, the discretization is performed by partitioning
the local frame into super-cells, and aggregating the
resolution elements belonging to each super-cell into 
the target property of the super-cell.
The partitioning scheme is required to be the same for all galaxies in 
the ensemble, so that the properties of the super-cells at the same 
local-frame coordinates across galaxies can be reduced further to construct 
the composite field.
Finally, the reduction step is performed 
to obtain the composite field ($f_E$) from those of individual
galaxies ($f_g$) as:
\begin{equation}
    f_E({\bm x}) \equiv \mathcal{R}_{g \in E} \left[ f_g({\bm x}) \right]\,,
    \label{eq:stacking}
\end{equation} 
where the reduction operator $\mathcal{R}$ is applied to the super-cells
at the same local-frame coordinate ${\bm x}$ from all galaxies in $E$. 
In this paper, we employ the median operator, rather 
than the mean, to robustly suppress irregularities from e.g. mergers, interactions 
and satellites that obscure the pattern of baryon cycle
underlying the ensemble. A more general formulation of the stacking method 
and the reduction operator is provided in Appendix~\ref{app:ssec:formulation}.

We can also compress the 3D field into lower-dimensional descriptors. 
This is achieved by extracting super-cells within a specified sub-volume 
of the local frame and reducing their properties to a single value. 
For example,
to obtain a spherically averaged profile $f_E(r)$ as a function of galactocentric 
distance $r$, we gather all super-cells within a spherical shell with a finite 
thickness $\Delta r$ centered at $r$ from all
galaxies in the ensemble, and apply the reduction operation to these super-cells 
to obtain the value of the composite field at $r$. 

\subsection{The ensembles of simulated galaxies}
\label{ssec:sample}

We apply the stacking method to the cosmological hydrodynamic simulation IllustrisTNG
\cite{nelsonFirstResultsTNG502019,pillepichFirstResultsTNG502019,nelsonIllustrisTNGSimulationsPublic2019}.
Our focus is on an ensemble of central star-forming galaxies with stellar mass 
$M_* \approx 10^{10} \Msun$ at redshift $z=0$. This selection is made to ensure 
that the galaxy ecosystems have similar spatial extents and comparable 
baryon contents, thereby enhancing the effectiveness of the stacking process. 
A total of 131 galaxies meet these criteria, with a median host-halo 
size of $R_{\rm h} \approx 149.0 \Kpc$. Other ensembles of galaxies with 
different $M_*$, $z$ and central supermassive black hole (SMBH) mass 
($M_{\rm BH}$) are also constructed to examine the dependence of the 
baryon cycle on these variables (see Appendix~\ref{app:ssec:sample} for the 
details of sample construction and Table~\ref{tab:subsamples} for a summary).

For our sample of star-forming galaxies, which are predominantly disk-like, 
the choice of local frame for the alignment step
is straightforward: we adopt a cylindrical frame with 
origin at the galaxy center, axis perpendicular to the disk plane 
(with positive direction specified by the gas angular momentum vector),
and an arbitrary azimuthal reference direction. 
The three coordinates of this frame are the galactocentric distance $r$,
the inclination angle $\theta$ relative to the disk plane, and the azimuthal angle $\phi$. 
Following ref.~\cite{wangFormationStarformingDisks2023}, we define the disk 
plane by minimizing the mass-weighted sum of distances from stellar 
particles to the plane. 

For each galaxy, we obtain fields for five gas properties: density ($n$, 
in units of mean nuclear weight), temperature ($T$), 
velocity vector [$\bm v$, including both the radial ($v_r$) and azimuthal 
($v_\phi$) components], the spin parameter (i.e. normalized specific angular momentum) 
along the disk normal direction ($\lambda$), and the flux density of the radial 
mass flow ($I$). The composite fields of these properties 
are then reduced from the aligned fields of individual galaxies in the 
ensemble.
The detailed definitions of these properties, as well as the 
fields around individual galaxies, are provided in 
Appendix~\ref{app:ssec:individual}.

\begin{figure*}[t]
	\centering
	\includegraphics[width=\textwidth]{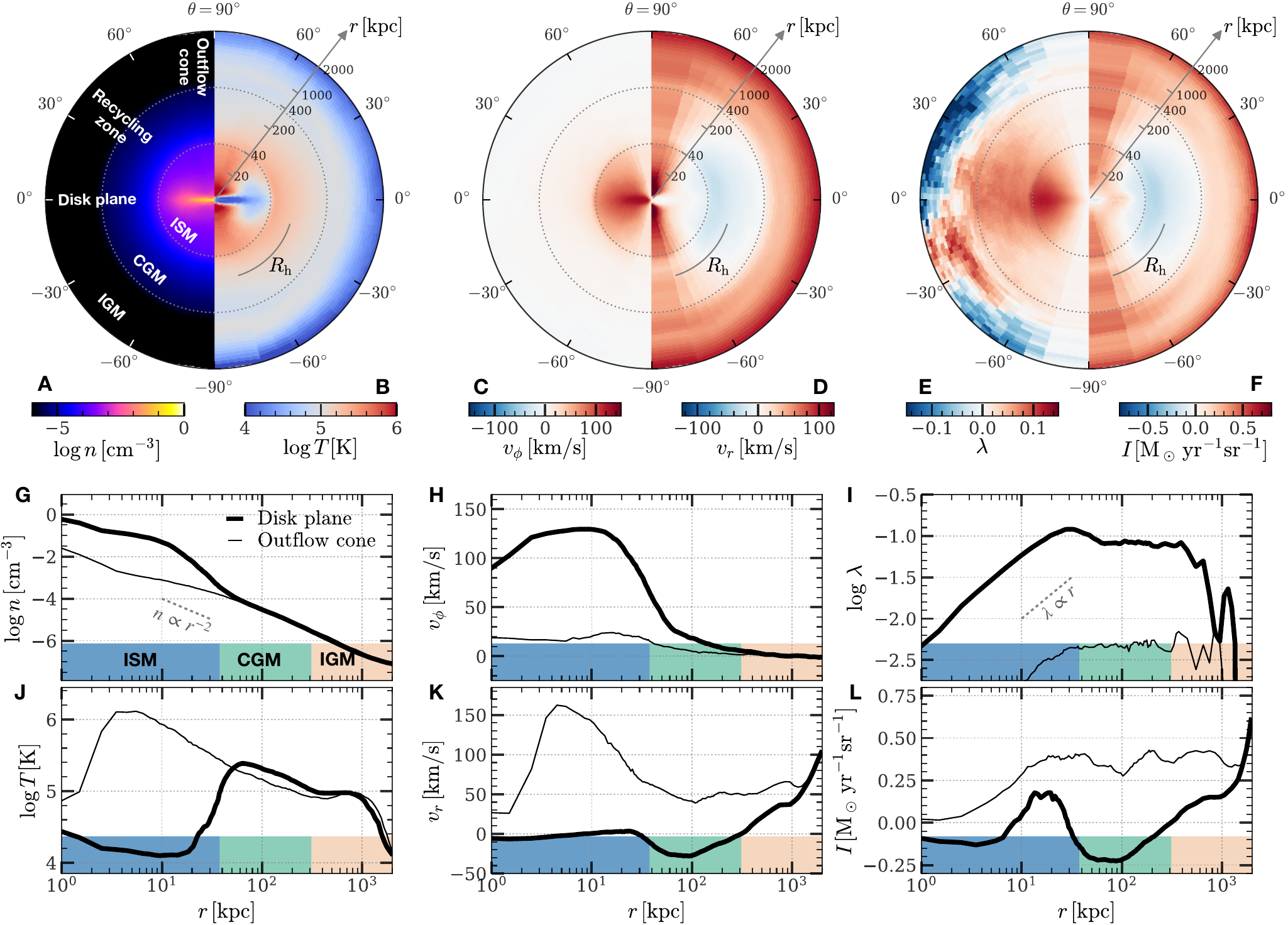}
    \caption{
        {\figem Profiles of gas properties in a stacked ensemble of galaxy ecosystems.}
        Here we show the fiducial ensemble (the same as that shown in Figure~\ref{fig:streamline}).
        {\figem (A)}--{\figem (F)} 2D maps showing gas density ($n$), temperature ($T$), 
        rotational velocity ($v_\phi$), 
        radial velocity ($v_r$), spin $\lambda$, and radial mass flux ($I$), 
        as functions of galactocentric distance $r$ 
        and inclination angle $\theta$ relative to the disk plane.
        Radial coordinates are shown by using piecewise-linear scaling within 
        three intervals, 
        $[0, R_{\rm ISM}]$, $[R_{\rm ISM}, R_{\rm CGM}]$ and 
        $[R_{\rm CGM}, 2\Mpc]$, 
        corresponding to the regimes of ISM, CGM and IGM, allowing visualization 
        across large dynamic ranges while preserving detail in each regime. 
        An arc in each panel marks the median halo virial radius 
        ($R_{\rm h}$).
        {\figem (G)}--{\figem (L)} Radial profiles along disk plane 
        ($\theta = 0$; thick curves) and outflow-cone center
        ($\left|\theta\right| = 90^\circ$; thin curves).
        Azimuthal angle ($\phi$) was integrated out to produce this figure.
        Distinct patterns of gas properties across the regimes validate our 
        divisions of galaxy ecosystems and reveal the underlying mechanics
        driving the fountain. 
        }
        \label{fig:mass10z0}
\end{figure*}

\begin{figure*}[t]
    \centering
    \includegraphics[width=0.9\textwidth]{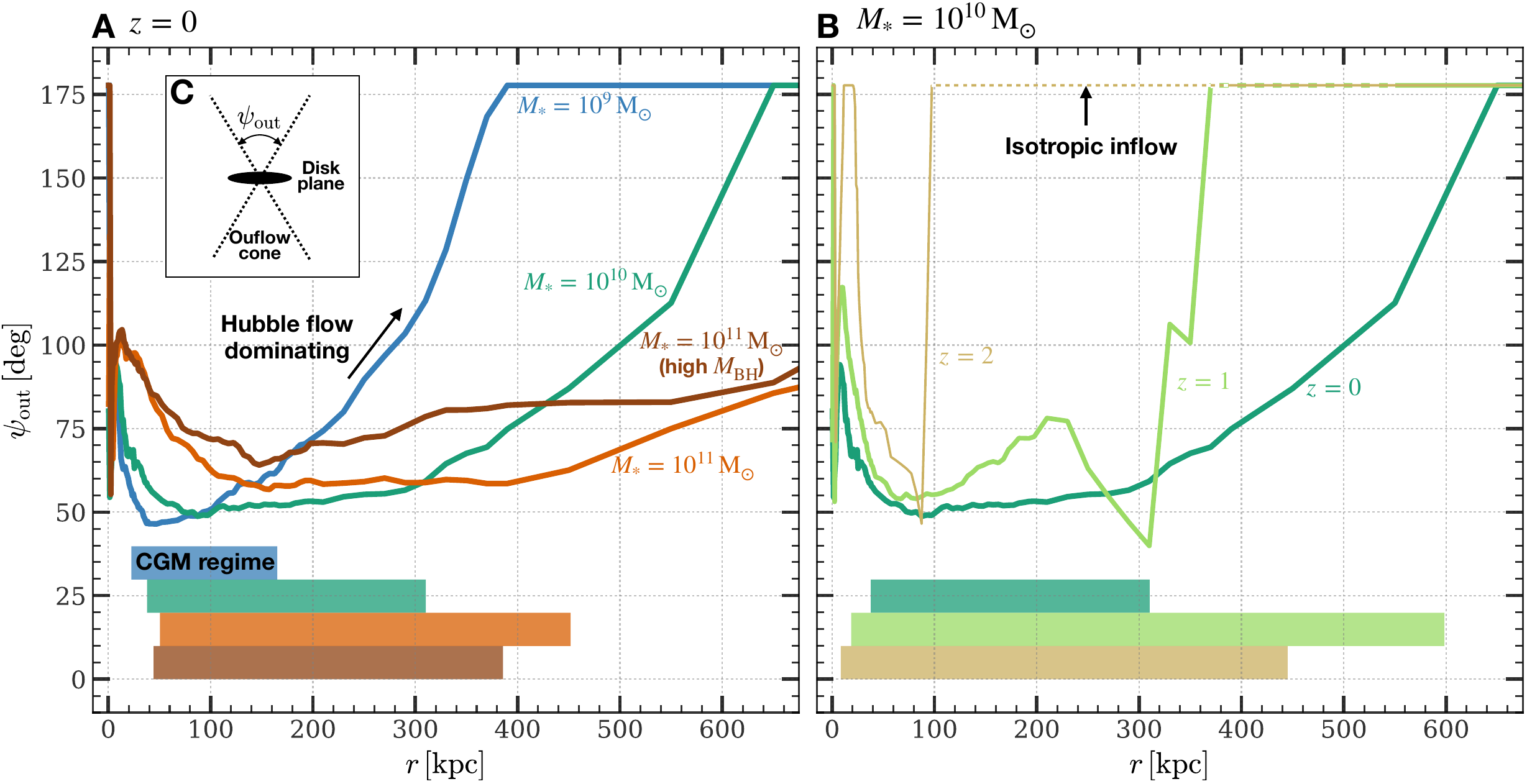}
    \caption{
    {\figem Opening angles of outflow cones. }
    {\figem (A)} Results for the ensembles with different $M_*$ and $M_{\rm BH}$ at $z=0$.
    {\figem (B)} Results for the ensembles at different $z$ with the same $M_*$ and $M_{\rm BH}$.
    {\figem (C)} Illustration of the definition of the opening angle 
    ($\psi_{\rm out}$).
    In {\figem (A)}, {\figem (B)}, colored bands indicate the CGM regimes 
    ($R_{\rm ISM} \leqslant r < R_{\rm CGM}$) for the ensembles. 
    Beyond the CGM regime, outflow gradually joins the Hubble flow.
    At $z=1$ and $z=2$, isotropic inflows dominate at intermediate radii
    in which $\psi_{\rm out}$ is not well defined and thus represented 
    as dashed segments.
    }
    \label{fig:outflow-geometry}
\end{figure*}

\begin{figure*}[!htbp]
	\centering
    \includegraphics[width=0.825\textwidth]{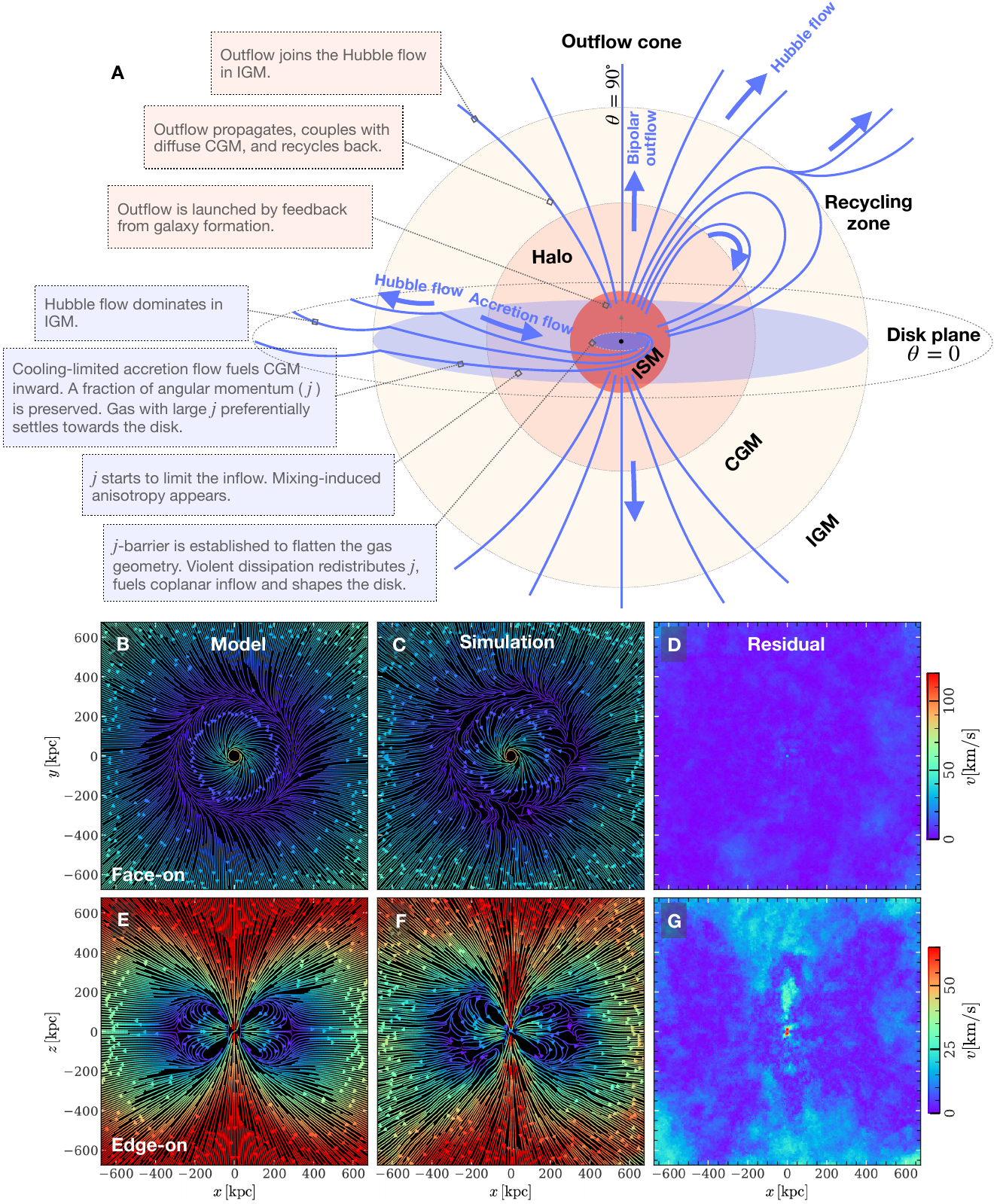}
    \caption{
        {\figem Parameterization of the fountain.}
        {\figem (A)} Schematic diagram (not to scale) summarizing the concepts defined to 
        describe galaxy ecosystems, and processes governing the baryon 
        cycle within the divisions. 
        The model is designed to encapsulate these concepts and processes,
        and the parameters specific to the IllustrisTNG ensemble are 
        obtained to match the stacked velocity field.
        {\figem (B)}--{\figem (D)} Face-on view of the velocity field 
        predicted by the model, obtained from the simulation (the same 
        as that shown in Figure~\ref{fig:streamline}A), and 
        the residual (norm of the vector difference) between 
        the two, for the fiducial ensemble.
        {\figem (E)}--{\figem (G)} Similar but for the edge-on view.
        The model can serve as a basis for compressing the physical understanding 
        extracted from the ecosystems while allowing the fountain pattern to be
        quantitatively restored upon demand.
    }
    \label{fig:model}
\end{figure*}

\section{Results}
\label{sec:results}

\subsection{Fountain pattern of baryon cycle}
\label{ssec:pattern}

Figure~\ref{fig:streamline} shows the stacked velocity field ($\bm v$) extracted 
from a slice that is either along (panels A--C; labeled ``face-on'') 
or perpendicular to (panels D--F; labeled ``edge-on'') the disk planes of the galaxies, 
with the velocity projected onto the slice. The field is depicted using 
streamlines, where arrows indicate the direction and colors represent the 
magnitude of the velocity. This visualization is presented in three zoom-in levels. 
In both slices, the streamlines extend from the galactic centers to scales 
well beyond the virial radii of the host halos, showing a smooth flow pattern 
in contrast to the chaotic pattern usually seen in individual galaxy 
ecosystems (see Appendix~\ref{app:ssec:individual} 
and Figure~\ref{fig:streamline-individual}; see also
refs.~\cite{springelCosmologicalSmoothedParticle2003,
pillepichSimulatingGalaxyFormation2018,
nelsonFirstResultsTNG502019,
marascoHIObservedSimulated2025}). 

The face-on view reveals a clear pattern as follows. On large scales 
(galactocentric distances $r \gtrsim 300 \Kpc$), gas is observed to move outward 
in all directions, reflecting cosmic expansion. On smaller scales ($r \lesssim 300 \Kpc$), 
in contrast, the gas flows inward due to gravitational attraction of the local mass 
concentration. As one approaches the galactic centers, rotational motion becomes increasingly dominant, 
indicating a transition from dispersion support to rotation support 
in the gas dynamics. 

More strikingly, the edge-on view reveals a ``{\texteem butterfly}'' 
pattern, showing the coexistence of gas inflow around the disk plane, 
a bipolar outflow perpendicular to the disk plane, and a recycling flow
in between. This smooth and coherent pattern reinforces our 
hypothesis that the stacking method effectively suppresses irregularities 
present in individual galaxy ecosystems, thereby revealing the underlying 
dynamics governing the ensemble.

\subsection{Ecosystem divisions and properties}
\label{ssec:divisions}

The clear pattern allowed us to rigorously define key concepts of galaxy 
ecosystems that have remained vague so far in the literature.
We define $R_{\rm CGM}$, the outer radius of the CGM regime, as the 
galactocentric distance where the disk-plane radial velocity $v_r$ changes 
sign (marked by the arc in Figure~\ref{fig:streamline}A),
mimicking the concept of ``turnaround'' in the definition of 
the halo boundary \cite{pavlidouWhereWorldStands2014,leeTurningCosmicWeb2016,
liOutermostEdgesMilky2021,hanManyBoundariesStratified2026}.
This definition marks the transition from the gravity-dominated inflow to the cosmic-expansion-dominated 
outflow, thereby specifying the volume within which the infall of matter 
directly participates in galaxy evolution. 
Matter beyond $R_{\rm CGM}$ resides in the IGM regime, where it interacts with galaxies 
only indirectly through, for example, gravity, radiation and outflows.
We define $R_{\rm ISM}$, the outer radius of the ISM regime, as the radius 
where the azimuthal rotational velocity $v_\phi$ drops to half of its maximum 
value as one moves outward.
This definition captures the transition from rotation-dominated to 
dispersion-dominated gas dynamics and, as discussed below,
is closely related to the separation between star-forming and
non-star-forming gas.
For our fiducial ensemble, these definitions yield $R_{\rm CGM} = 310.3\,\Kpc$ and 
$R_{\rm ISM} = 37.8\,\Kpc$ (see Table~\ref{tab:subsamples}).
We can further subdivide galaxy ecosystems by the angular geometry 
into a cone around the polar axis (with inclination angle $\left|\theta\right| = 90^\circ$), 
the disk plane ($\theta = 0$) and the recycling zone between the two.

To see the distinct properties of each ecosystem division, we follow the method in \S\ref{ssec:stacking-method} to reduce the stacked fields of gas properties into 2D maps by integrating over the azimuthal angle (Figure~\ref{fig:mass10z0}A--F), and further compress them into radial profiles along the disk plane ($\theta = 0$) and the outflow-cone center ($\left|\theta\right| = 90^\circ$; Figure~\ref{fig:mass10z0}G--L).
The ISM regime exhibits a dense, cold gaseous disk clearly visible in the 2D maps 
of density and temperature. Its thin geometry demonstrates that the stacking method 
successfully aligns disks across individual galaxies. The CGM regime contains diffuse, 
heated gas with low rotational velocity and detectable inflow and outflow. 
At large radii, 
the disk-plane profiles of density and temperature closely resemble those in the 
outflow cone; however, they diverge sharply at $R_{\rm ISM}$, producing the strong 
anisotropy characteristic of the ISM (see also Figure~\ref{fig:mass10z0-angular})
and raising gas density to the level required to trigger star formation \cite{pillepichSimulatingGalaxyFormation2018}. 
The disk-plane spin profile exhibits notable structure across the regimes: 
it is noisy in the IGM, indicating a lack of coherent large-scale rotation aligned 
with the disk; it stabilizes into a plateau in the CGM, peaks at $r\approx R_{\rm ISM}$, 
and decreases nearly linearly toward galactic centers. 
The sector-like pattern in the 2D spin map underscores the critical role of 
angular momentum in flattening the ISM geometry and shaping the 
disks (see Appendix~\ref{app:ssec:discussion-am} for further discussion; 
see also \cite{moFormationGalacticDiscs1998,wangOriginExponentialStarforming2022}).
The disk-plane radial velocity shows a net inflow that accelerates in the CGM 
and decelerates at the ISM-CGM interface, accompanied by shock heating, as 
is evident in the temperature upturn at $r \lesssim R_{\rm CGM}$. In contrast, the 
velocity field around the polar axis displays net outflow that is launched within the ISM 
and remains roughly constant through the CGM and IGM. 
These distinct features within each division of the ecosystems validate our physical 
definitions of the divisions and confirm that our stacking method effectively reveals the 
underlying pattern governing the galaxy ensemble.
More details on the ecosystem divisions and properties are provided in 
Appendix~\ref{app:ssec:field-ensemble}.

\subsection{Outflow geometry}
\label{ssec:outflow-geometry}

The smooth pattern in the stacked velocity field enables quantitative 
characterization of outflow geometry, which was infeasible for individual galaxy 
ecosystems due to their irregularity (see Figure~\ref{fig:streamline-individual}
for examples of individual ecosystems).
From the 2D map of $v_r$ (Figure~\ref{fig:mass10z0}D), 
we extract the angular profiles of radial velocity ($v_r$ versus $\theta$) 
at different $r$. At each $r$, we identify the boundaries of the 
outflow cone as the locations of $\theta$ at which $v_r$ drops to half of the
value at $\left|\theta\right|=90^\circ$, and we define the opening angle ($\psi_{\rm out}$)
of the outflow cone as the angle enclosed by these boundaries 
(see Figure~\ref{fig:outflow-geometry}C for an illustration). 
$\psi_{\rm out}$ so obtained for the fiducial ensemble is shown 
in Figure~\ref{fig:outflow-geometry}(A), which varies with $r$ as outflow propagates 
outward through the ecosystems.
$\psi_{\rm out}$ fluctuates with $r$ in the ISM regime where the 
outflow is launched; out to the CGM regime, $\psi_{\rm out}$ converges to
a nearly constant value of $\approx 50^\circ$--$60^\circ$.
This, together with the flat $v_r$ and $I$ profiles in the outflow 
cone at CGM scales (see Figure~\ref{fig:mass10z0}K,L), indicates that 
the outflow maintains a collimated structure throughout the CGM, as already 
evident from the streamlines (see Figure~\ref{fig:streamline}, edge-on view).
At $r \gtrsim R_{\rm CGM}$, the outflow gradually merges into the isotropic 
Hubble flow, as seen from the increase of $\psi_{\rm out}$ toward 
$180^\circ$.

\subsection{Formation of fountain in $\Lambda$CDM paradigm}
\label{ssec:fountain-in-lcdm-main}

The fountain pattern, ecosystem divisions, and quantitative properties of gas 
within each regime can be understood within the $\Lambda$CDM structure-formation 
paradigm \cite{peeblesLargescaleStructureUniverse1980,moGalaxyFormationEvolution2010}. 
We present a thorough discussion of the formation of the fountain in 
Appendix~\ref{app:ssec:fountain-in-lcdm}; here we summarize the key points.
The inflow generated in the 
outer CGM is consistent with predictions from virialized, cooling-limited 
steady-state models. The plateau-like profile of the spin in the CGM indicates that a 
substantial fraction of the angular momentum is preserved during the inflow of the gas. 
The peak in the disk-plane profile of spin at $r \approx R_{\rm ISM}$ 
reflects violent angular-momentum redistribution at the ISM-CGM interface, which
allows some gas elements to overcome the angular-momentum barrier
and to be accreted into the ISM \cite{goldnerAccretionDrivenTurbulenceCircumgalactic2025,hobbsFeedingSupermassiveBlack2011,gaspariChaoticColdAccretion2015}. 
The pattern of the velocity field in the ISM (see Figure~\ref{fig:streamline}F) 
and the declining disk-plane inflow flux towards galactic centers (see Figure~\ref{fig:mass10z0}L) suggest that gas 
acquisition follows a coplanar scenario \cite{lyuDominantRoleCoplanar2025} in and around gaseous disks. 
The bipolar geometry of the outflow is not a result of the isotropic injection of
feedback \cite{pillepichSimulatingGalaxyFormation2018,
weinbergerSimulatingGalaxyFormation2017},
but is a direct consequence of the disk structure of the ISM, which 
creates a preferred pathway of low resistance for the feedback-driven gas to 
flow out \cite{springelCosmologicalSmoothedParticle2003,
sivasankaranAGNFeedbackIsolated2025}. 
The smooth, nearly isotropic radial profiles of density 
and temperature in the CGM indicate that outflows constitute 
a subdominant mass component relative to the ambient gas and
heat the CGM only mildly in galaxy ecosystems 
(see \S3.4 of ref.~\cite{nelsonFirstResultsTNG502019}). 
The large spatial extent and collimated geometry of the outflow into the CGM and IGM reflect 
the substantial energy injection from the feedback of the central galaxy  
and the low density of the diffuse CGM and IGM environment that provides little resistance 
to the propagation of the outflow.

\subsection{Parameterizing the fountain pattern}
\label{ssec:model-main}

The quantification and understanding of galaxy ecosystems allow us to develop a parametric 
model that describes the processes driving the baryon cycle as a set of equations,
with parameters calibrated to match the stacked fields for a specific ensemble of 
galaxies (see Figure~\ref{fig:model}A for a schematic summary,
and Appendix~\ref{app:ssec:model} for details of the parameterization and calibration). 
We calibrated the model 
using the stacked velocity field of the fiducial ensemble, resulting in a predicted 
velocity field (Figure~\ref{fig:model}B,E). The model recovers the fountain pattern 
revealed in the simulation (Figure~\ref{fig:model}C,F), with moderate 
residuals (Figure~\ref{fig:model}D,G). The calibrated parameters possess explicit 
physical meanings (see Appendix~\ref{app:ssec:model}), indicating that the 
imprints of the processes shaping the fountain can be neatly encoded. 
The model thus serves as a basis for compressing the physical understanding 
extracted from the ecosystems while allowing the fountain pattern to be
quantitatively restored upon demand.

\subsection{Variations across galaxy ensembles}
\label{ssec:variations}

In addition to the fiducial ensemble presented above,
we also apply the stacking method to ensembles of galaxies with varying $M_*$, $z$, 
and $M_{\rm BH}$. We present quantitative results and detailed discussions 
in Appendices~\ref{app:ssec:mass-dependence}, 
\S\ref{app:ssec:redshift-dependence} and \S\ref{app:ssec:discussion-bh}, respectively,
and summarize the key findings here.
We find fountain patterns in the gas flow similar to those found for the fiducial ensemble. 
At fixed redshift, the physical extent ($R_{\rm ISM}$ and $R_{\rm CGM}$) of the fountain 
scales with halo size ($R_{\rm h}$); 
galaxies in larger halos develop more extended ISM regimes and outflow coverages, 
highlighting the fundamental role of dark matter halos in shaping galaxy 
ecosystems \cite{wechslerConnectionGalaxiesTheir2018,jiangDarkmatterHaloSpin2019,lyuHalosGalaxiesVII2023}. 
These findings align with empirical constraints from observations. 
Systematic trends also emerge 
across ensembles of galaxies at different redshifts. Higher-redshift galaxies exhibit hotter 
gas dynamics, more isotropic inflows and outflows, and larger inflow and 
outflow fluxes, indicating that steady ecosystems have yet to be established 
due to the rapid growth of structures in the early Universe. 
Galaxies hosting massive SMBHs are embedded in ecosystems with more compact
configurations and lower gas spin, suggesting an interplay among structures across
multiple scales \cite{zjupaAngularMomentumProperties2017,
rodriguez-gomezRoleMergersHalo2017,maEvolutionaryPathwaysDisk2024,
duPhysicalOriginMasssize2024,wuOvermassiveBlackHoles2025}.
All these differences are naturally explained within the $\Lambda$CDM paradigm, 
indicating that a reliable framework 
can be constructed to describe and understand galaxy ecosystems within the 
current scenario of structure formation. 

The uniformity of the fountain pattern and the variations in the detailed 
structure across ensembles are further confirmed by examining the outflow geometry, 
as quantified by $\psi_{\rm out}$ (Figure~\ref{fig:outflow-geometry}A,B).
The remarkable consistency of $\psi_{\rm out}$ in the CGM is seen across all ensembles 
at $z=0$, with more massive galaxies exhibiting slightly wider 
angles. 
At higher redshifts, the range of radii where $\psi_{\rm out}$ remains small 
is more limited, consistent with the more isotropic outflow geometry 
seen in the 2D maps (Figure~\ref{fig:zdependence}).
At $z=1$ and $z=2$, the outflow diminishes at large radii and mixes into 
the isotropic inflow (dashed segments in Figure~\ref{fig:outflow-geometry}B)
in the CGM regime. 
The presence of such an outflow horizon is a consequence of 
the limited velocity and time for the outflow to propagate against the 
inflow that becomes increasingly strong with lookback time owing to the rapid 
growth of halos in the early Universe.
This horizon marks the maximal radius within which outflow dominates the
chemical composition of gas, and therefore defines an expanding zone of 
influence within which star formation in nearby mini-halos is gradually 
modified \cite{venturaSemianalyticModellingPop2024,chenTwophaseModelGalaxy2025a}.
The build-up of a steady fountain is thus a gradual process associated with disk 
settling and stabilization of the halo potential, and baryon cycles in the 
early Universe proceed in a remarkably different and more rapidly evolving manner,
compared with those in ecosystems with similar $M_*$ at low redshift.

\section{Discussion and conclusion}
\label{sec:summary}

In this paper, we have developed a stacking method to produce composite fields around ensembles of galaxies, so that irregularities in individual galaxy ecosystems are suppressed and the underlying pattern of baryon 
cycle can be recovered (\S\ref{sec:method}).
Applying this method to star-forming galaxies in the IllustrisTNG
simulation, we have revealed a strikingly regular fountain pattern of 
baryon cycle. This is in contrast to the chaotic patterns seen in 
individual ecosystems, indicating that the fountain pattern is an ensemble feature, 
rather than a property of individuals
(\S\ref{ssec:pattern}; Figure~\ref{fig:streamline}).
The regular pattern allows us to define key concepts 
of galaxy ecosystems, including the ISM, CGM, and IGM regimes, 
to quantify the properties of baryons within each regime, 
and to extract descriptors such as the opening angle of 
the outflow cone (\S\S\ref{ssec:divisions}, \ref{ssec:outflow-geometry};
Figures~\ref{fig:mass10z0}, \ref{fig:outflow-geometry}).
The established fountain pattern can be understood within the $\Lambda$CDM structure-formation paradigm,
and can be encoded as a set of equations with 
informative parameters (\S\S\ref{ssec:fountain-in-lcdm-main},
\ref{ssec:model-main}; Figure~\ref{fig:model}).

The stacking method is not limited to our fiducial ensemble,
since none of its steps relies on simulation-specific 
implementations:
it is applied to ensembles of star-forming galaxies with 
different $M_*$, $z$, and $M_{\rm BH}$, all of which exhibit 
similar but quantitatively different fountain patterns
(\S\ref{ssec:variations});
it is straightforward to adapt the method to other galaxy 
populations and galaxies in simulations with different 
prescriptions. We note that the optimal schemes for applying the stacking method
to other galaxy populations and galaxies in other simulations, 
for decomposing their ecosystems into divisions,
for interpreting the revealed patterns within the structure-formation paradigm, and for comparing the results between simulations and observations, may differ from those adopted here.
Our method thus provides a basis to explore such variations.

The fountain pattern revealed in IllustrisTNG is
consistent with that established by combining pieces of evidence from 
observations with theoretical understanding (see Supplemental Figure~1 and \S3 
of ref.~\cite{tumlinsonCircumgalacticMedium2017}), 
supporting the use of cosmological hydrodynamic simulations to 
recover the baryon cycle and to interpret observational data. 
Meanwhile,
the fountain pattern revealed here can inform observational strategies aimed at
mapping gas surrounding galaxies. Notably, the bipolar geometry of outflow suggests that 
heavy elements yielded in the inner ecosystems, when transported outward, should 
exhibit a similar bipolar distribution. This bipolar enrichment is indeed found in IllustrisTNG 
(see Appendix~\ref{app:ssec:metal} and Figure~\ref{fig:metal-field}). 
Observationally, stacking has proven valuable for detecting weak signals 
in ensembles of galaxy ecosystems \cite{limGasContentsGalaxy2018,lanCircumgalacticMediumEBOSS2018,wuConstraintsCircumgalacticMedia2020,
popessoXrayInvisibleUniverse2024,zhangHotCircumgalacticMedium2024,chenCircumgalacticMediumTraced2025}, 
and may thus be applied to trace the distribution of metals, to constrain
the degrees of freedom of the parametric model, and to recover the underlying 
fountain pattern.
Our stacking approach for simulated galaxies, combined with rigorous ecosystem 
divisions, quantitative characterization of the baryonic field in each division, 
and parametric modeling, offers a physically motivated, 
yet intuitive framework for linking simulations with observations.

\Acknowledgements{This work is supported by the National Natural Science Foundation of China (Grant nos. 12473008 and 12503014), 
the Fundamental Research Funds for the Central Universities (Grant no. KG202502) and the Double First-Class Discipline Construction-Astronomy Discipline at Nanjing University.
The authors thank Meicun Hou and Houzun Chen for comments, and express their 
gratitude to the Tsinghua Astrophysics 
High-Performance Computing platform at Tsinghua University for 
providing the necessary computational and data storage resources that have 
significantly contributed to the research results presented in this paper.}

\InterestConflict{The authors declare that they have no conflict of interest.}


\bibliographystyle{scpma} 
\bibliography{ref}

\begin{appendix}




\renewcommand{\thesection}{Appendix}
\setcounter{figure}{0}

\section{}

\subsection{\label{app:ssec:sample}Sample selection} 

In this paper, we analyzed the cosmological hydrodynamic simulation
conducted as part of the IllustrisTNG project \cite{nelsonIllustrisTNGSimulationsPublic2019,
springelFirstResultsIllustrisTNG2018,pillepichFirstResultsIllustrisTNG2018,
nelsonFirstResultsIllustrisTNG2018,
marinacciFirstResultsIllustrisTNG2018}.
Specifically, we used TNG50-1, the run with the highest resolution among those publicly available, 
to minimize numerical artifacts when analyzing baryonic properties.
We defined halo properties as follows. The virial mass ($M_{\rm h}$) is the total 
(dark and baryonic) mass of matter within the virial radius ($R_{\rm h}$), 
defined as that enclosing a mean matter density of $\Delta_{\rm v} \equiv 200$ 
times the critical density of the Universe \cite{navarroUniversalDensityProfile1997}. 
The virial velocity ($V_{\rm h}$) is the circular velocity at $R_{\rm h}$. 
The aperture used to evaluate the properties of a galaxy is twice 
the stellar half-mass radius. Distances displayed in this paper 
are in physical units.

Various processes can drive gas out of galaxies, including 
stellar feedback \cite{oppenheimerFeedbackRecycledWind2010,muratovGustyGaseousFlows2015,yuStellarFeedbackDrives2025,lyuFirstStatisticalDetection2026}, 
active-galactic-nucleus (AGN) feedback \cite{sivasankaranAGNFeedbackIsolated2025,deugenioFastrotatorPoststarburstGalaxy2024},
and environmental effects such as 
tidal stripping \cite{arrakiEffectsBaryonRemoval2014,wangEvolutionaryContinuumNucleated2023} 
and ram pressure stripping \cite{wangRamPressureStripping2020,liShockinducedStrippingSatellite2023}. 
To focus on the underlying baryon cycle without complications from quenching mechanisms, 
we restricted our main analyses to central galaxies on the star-forming main sequence (SFMS).
We defined the SFMS using an iterative approach \cite{wooDependenceGalaxyQuenching2013,
donnariStarFormationActivity2019,wangDissectTwohaloGalactic2023,maRevisitingFundamentalMetallicity2024}. 
Starting from all central galaxies with $M_* \geqslant 10^8 \Msun$ at a given redshift $z$, 
we fitted the star formation rate (SFR)-$M_*$ relation in logarithmic space:
\begin{equation}
    \log\left(\frac{\rm SFR}{\rm M_\odot yr^{-1}}\right) = a\,\log\left(\frac{M_*}{\rm M_\odot}\right) + b
    \,,\label{eq:SFMS-fit}
\end{equation}
where $(a,b)$ are free parameters. We identified star-forming galaxies as those 
above $-1\dex$ of this relation and refitted the relation. The fitting and selection
were iterated until convergence. The best-fit parameters at $z=0$, $1$ and $2$ are 
$(a,b) = (0.748, -7.708)$, $(0.970, -9.199)$ and $(1.071, -9.708)$, 
respectively.
Since AGN feedback quenches IllustrisTNG galaxies primarily through 
kinetic-mode feedback, and this process is not instantaneous \cite{liPhysicalProcessesCoevolution2025,
terrazasRelationshipBlackHole2020,wardCosmologicalSimulationsPredict2022,jiangDissectingMassQuenching2025}, we additionally excluded galaxies 
with $M_{\rm BH} > 10^{8.2}\Msun$ to ensure kinetic-mode AGN feedback does not significantly
alter the baryon cycle in our fiducial ensemble \cite{jiangDissectingMassQuenching2025}. 
The fraction of galaxies excluded by this $M_{\rm BH}$-criterion increases with $M_*$
(about $1\%$ among star-forming galaxies at $M_* \approx 10^{10} \Msun$ at $z=0$, and 
about $66\%$ at $M_* \approx 10^{11} \Msun$).
The ensemble with high $M_{\rm BH}$ is discussed in Appendix~\ref{app:ssec:discussion-bh}.

From the galaxies meeting the above selection criteria, we constructed 
ensembles by restricting the ranges of stellar mass and redshift. 
This approach was motivated by observations that $M_*$ and $z$ are the 
primary determinants of galaxy size. When both quantities are fixed, the 
scatter in galaxy size is reduced to approximately $0.3\dex$ across cosmic 
history, as demonstrated by studies of the stellar size-stellar mass 
relation \cite{shenSizeDistributionGalaxies2003,somervilleRelationshipGalaxyDark2018,
nedkovaExtendingEvolutionStellar2021,jiaSizeGrowthShort2024,onoMorphologicalDemographicsGalaxies2025}. 
Combined with the tight observed correlation between stellar and neutral-hydrogen 
sizes \cite{panRoleRegulatingSize2021}, and the inferred correlation between 
stellar and halo sizes \cite{moFormationGalacticDiscs1998,
kravtsovSizeVirialRadiusRelation2013,
huangRelationsSizesGalaxies2017,
mishraStellarMassDependence2023}, 
galaxies within a fixed $M_*$-$z$ bin exhibit similar spatial extents of 
their baryon distributions. This ensures that such ensembles are
well-suited for stacking analyses without the need for coordinate 
rescaling. Conversely, mixing galaxies with very different 
spatial extents would cause the stacked field at any location to reflect 
physical processes operating across disparate scales, compromising 
physical interpretability. Table~\ref{tab:subsamples} 
summarizes the ensembles analyzed in this study. 
Ensemble 2 ($M_* = 10^{10}\Msun$, $z = 0$) serves as the fiducial one 
demonstrated in this paper.

\begin{table}[H]
\centering
\footnotesize
\begin{threeparttable}\caption{Ensembles of IllustrisTNG galaxies analyzed in this study}\label{tab:subsamples}
	\doublerulesep 0.1pt \tabcolsep 8pt 
	\begin{tabular}{c c c c c c}
	\toprule
		(1)
		&
		(2)
		&
		(3)
		&
		(4)
		&
		(5)
		&
		(6)
	\\
	\hline
		ID
		& $z$
		& $M_*$
		& $M_{\rm BH}$
		& $N_{\rm galaxies}$
		& $R_{\rm h}$, $R_{\rm ISM}$, $R_{\rm CGM}$ 
	\\
		& 
		& $[{\rm M}_\odot]$ 
		& $[{\rm M}_\odot]$ 
		&
		& $[{\rm kpc}]$ 
	\\
	\hline
		1 & $0$ & $ 10^{9\pm 0.05}$ &                         & $166$ & $97.1,\, 22.6,\, 165.2$ \\
		2 & $0$ & $ 10^{10\pm 0.1}$ &                         & $131$ & $149.0,\, 37.8,\, 310.3$ \\
		3 & $0$ & $ 10^{11\pm 0.4}$ & $\leqslant 10^{8.2}$    & $36$  & $234.7,\, 50.6,\, 451.7$ \\
		4 & $1$ & $ 10^{10\pm 0.1}$ &                         & $128$ & $108.2,\, 18.7,\, 598.3$ \\
		5 & $2$ & $ 10^{10\pm 0.1}$ &                         & $97$  & $82.1,\, 8.6,\, 445.1$ \\
	\hline
		6 & $0$ & $ 10^{11\pm 0.4}$ & $> 10^{8.2}$            & $69$  & $291.4,\, 44.3,\, 385.7$ \\
	\bottomrule
	\end{tabular}
	\begin{tablenotes}[flushleft]
		\item {\figem Columns}: 
		{\figem (1)} ensemble ID; 
		{\figem (2)} redshift; 
		{\figem (3)} stellar mass range; 
		{\figem (4)} SMBH mass range; 
		{\figem (5)} number of galaxies; 
		{\figem (6)} median halo virial radius $R_{\rm h}$, ISM outer radius $R_{\rm ISM}$, and CGM outer radius $R_{\rm CGM}$.
		\item {\figem Ensembles}:
		Ensemble 2 (fiducial) is used throughout the main text to demonstrate the 
		stacking method and to characterize the baryon cycle.
		Ensembles 1--3 span two decades in $M_*$ at fixed $z=0$, and are used in
		the analyses of stellar-mass dependence.
		Ensembles 2, 4 and 5 share the same $M_*$ but span $z = 0$ to $2$, 
		and are used in the analyses of redshift dependence.
		Ensembles 3 and 6 share the same $M_*$ and $z$ but differ in $M_{\rm BH}$,
		and are used in the analyses of SMBH-mass dependence.
		See Appendix~\ref{app:ssec:sample} for the details of sample construction.
\end{tablenotes}\end{threeparttable}
\end{table}

\subsection{\label{app:ssec:individual}Baryon distributions in individual galaxy ecosystems}

\begin{figure*}[t]
    \centering
    \includegraphics[width=\textwidth]{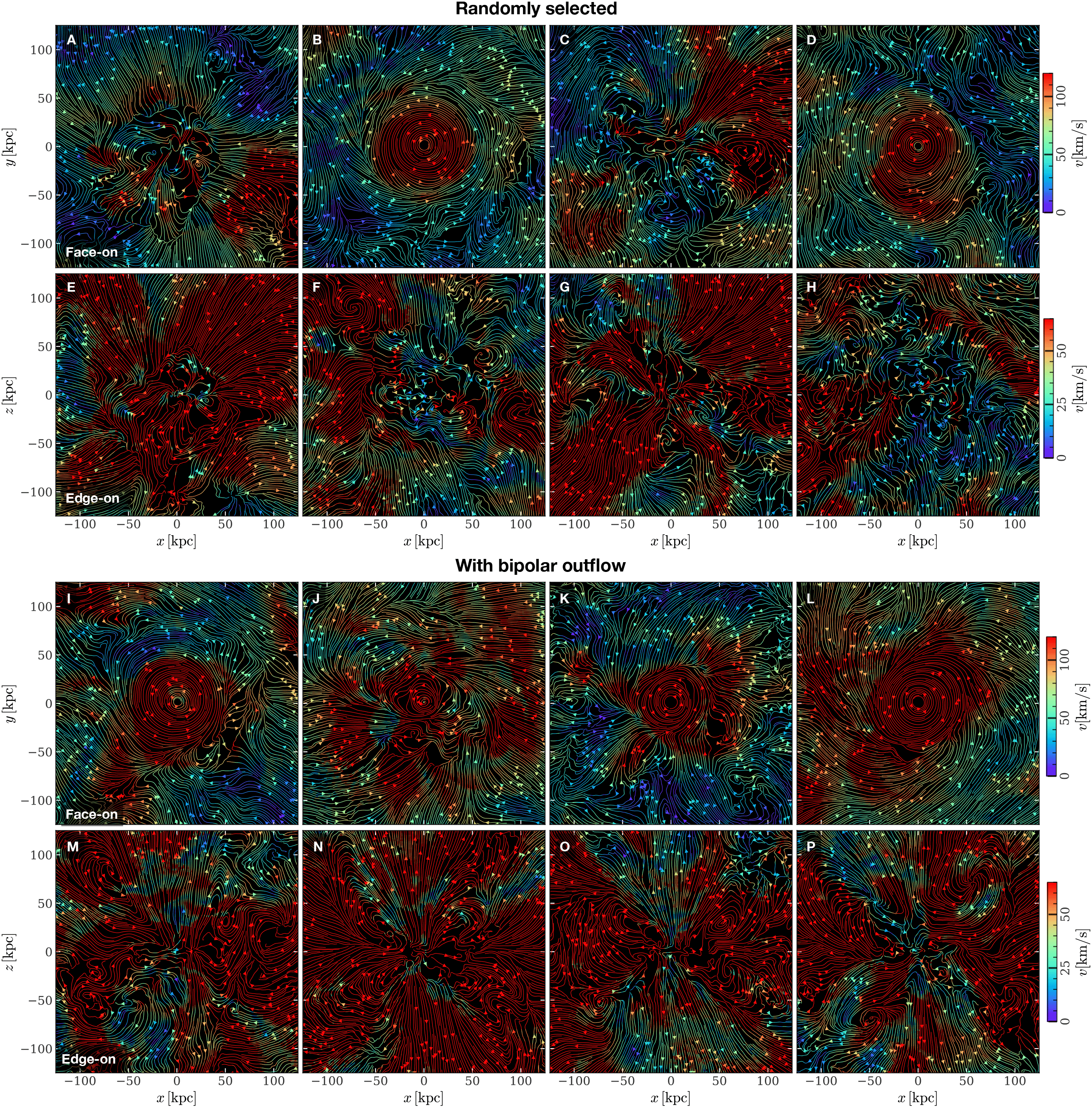}
    \caption{
    {\figem Velocity fields around individual galaxies.} 
    In the upper two rows, we show the velocity fields around four galaxies 
    randomly selected from 
    the fiducial ensemble (Table~\ref{tab:subsamples}, ID=2).
    {\figem (A)}--{\figem (D)} Face-on view for a slice along the disk plane
    of each galaxy.
    {\figem (E)}--{\figem (H)} Edge-on view for a slice perpendicular to the disk 
    plane of each galaxy.
    The fields are produced in the same way as Figure~\ref{fig:streamline},
    except that no stacking is applied here.
    {\figem (I)}--{\figem (P)}
    Similar but for four galaxies selected to exhibit prominent bipolar outflows.
    }
    \label{fig:streamline-individual}
\end{figure*}

We quantify the distribution of baryons around each simulated galaxy by 
defining a set of fields, each representing one baryonic property as a function 
of spatial location. As introduced in \S\ref{ssec:stacking-method}, 
to unify analyses across different simulations, 
we discretize the volume surrounding each galaxy into ``super-cells'', 
which are computational units large enough to contain sufficient resolution 
elements for robust evaluation of physical properties, yet small enough to 
capture spatial variations in the galaxy ecosystem. 
For each super-cell $\mathcal{C}$, 
we identify all resolution elements within it and aggregate 
their properties into a set of derived quantities. We analyze several gas properties
in this paper. The first two are gas density and temperature, 
defined as mass-weighted averages:
\begin{align}
    n &\equiv \frac{\sum_i m_i}{\mu_{\rm n} m_{\rm p} \sum_i m_i/\rho_i}
    \,, \label{eq:n-def} \\
    T &\equiv \frac{\sum_i m_i T_i}{\sum_i m_i}
    \,, \label{eq:T-def}
\end{align}
where the density is expressed as the number of nuclei per unit volume. 
Here, $m_i$ and $\rho_i$
are the mass and density, respectively, of the $i$-th gas cell
in $\mathcal{C}$; 
$\mu_{\rm n} = 1.22$ is the mean nuclear weight \cite{klessenFormationFirstStars2019}; 
$m_{\rm p}$ is the proton mass; 
and $T_i = (\gamma - 1)\mu_i m_{\rm p} u_i/k_{\rm B}$ is the effective temperature of 
the gas cell, with $\gamma=5/3$ and $k_{\rm B}$ the Boltzmann constant;
$u_i$ and $\mu_i$ are the specific internal energy and mean molecular weight
(with free electrons taken into account) of the gas cell. 
Note that density and temperature are fundamental thermodynamic variables 
that encode the signatures of physical processes regulating galaxy formation \cite{martizziBaryonsCosmicWeb2019,
torreyEvolutionMassmetallicityRelation2019}.
The next variable is the velocity vector, used to characterize gas kinematics:
\begin{equation}
    {\bm v} \equiv \frac{\sum_i m_i {\bm v}_{i}}{\sum_i m_i}
    \,, \label{eq:v-def}
\end{equation}
where ${\bm v}_i$ is the physical velocity (peculiar velocity plus the Hubble flow) of the $i$-th 
gas cell. We extract two key components: the radial velocity $v_r$ (indicating net inflow/outflow speed) 
and the azimuthal velocity $v_\phi$ (capturing rotational support, essential for our 
disk-dominated sample; see below 
for the definitions of disk plane and galactocentric cylindrical frame).
The final two, motivated by conservation laws, are spin 
(normalized specific angular momentum along the disk normal direction) 
and mass flux (radial mass flow rate per unit solid angle):
\begin{align}
    \lambda  &\equiv \frac{j}{\sqrt{2}R_{\rm h} V_{\rm h}}
    = \frac{\sum_i m_i v_{\phi, i} r_{{\rm p}, i}}{\sqrt{2} R_{\rm h} V_{\rm h} \sum_i m_i}
    \,, \label{eq:lambda-def} \\
    I  &\equiv \frac{\mathrm{d}\Phi}{\mathrm{d}\Omega} 
    = \frac{\sum_i m_i v_{r,i} r_i^2}{\sum_i m_i/\rho_i}
    \,,
\end{align}
where $j$ is the specific angular momentum along the disk normal direction;
$v_{\phi, i}$ and $v_{r,i}$ are the azimuthal and radial velocities 
of the $i$-th gas cell; $r_{{\rm p}, i}$ is the perpendicular distance 
from the cell to the cylindrical axis of the galactocentric cylindrical frame 
and $r_i$ is the galactocentric distance. 
Together, these properties provide a comprehensive characterization of the baryon 
cycle across galaxy ecosystems. Additional properties, as well as the 
weighting scheme among resolution elements, can be defined according to the 
goals of specific applications.

In Figure~\ref{fig:streamline-individual}(A)--(H), we show the velocity fields 
around four example galaxies randomly selected from the fiducial ensemble. 
Two galaxies display regular rotational motion in the CGM regime (panels B,D), 
whereas the other two exhibit less organized CGM kinematics. The bipolar 
signature of outflow is visible in two galaxies (panels G,H), albeit with
patterns deviating from cone-like shapes, and remains absent in the 
remaining two. Off-center attractors of streamlines indicate the presence of satellite
galaxies. These irregularities in individual 
galaxy ecosystems motivated the use of the stacking method to uncover the 
coherent patterns underlying the ensemble.
In Figure~\ref{fig:streamline-individual}(I)--(P), we show the velocity
fields around another four example galaxies from the same ensemble, selected to
have obvious bipolar outflow. The CGM of all four galaxies 
exhibits ordered rotation (panels I--L), indicating that coherent CGM kinematics
may be a prerequisite for the development of cone-like outflow geometry.
A comparison between the composite field (Figure~\ref{fig:streamline}) and those 
of individual galaxies demonstrates that the fountain pattern commonly described 
in the literature \cite{tumlinsonCircumgalacticMedium2017,wangOriginExponentialStarforming2022,
wangGasphaseMetallicityProfiles2022,churchillSpatialKinematicAbsorptionModels2025} 
does not typically 
manifest in most individual systems. Instead, this pattern should be interpreted 
as a statistical property that emerges when considering ensembles of galaxies.

\subsection{\label{app:ssec:field-ensemble}The stacked fields around ensembles of galaxies}

\begin{figure*}[t]
   \centering
    \includegraphics[width=\textwidth]{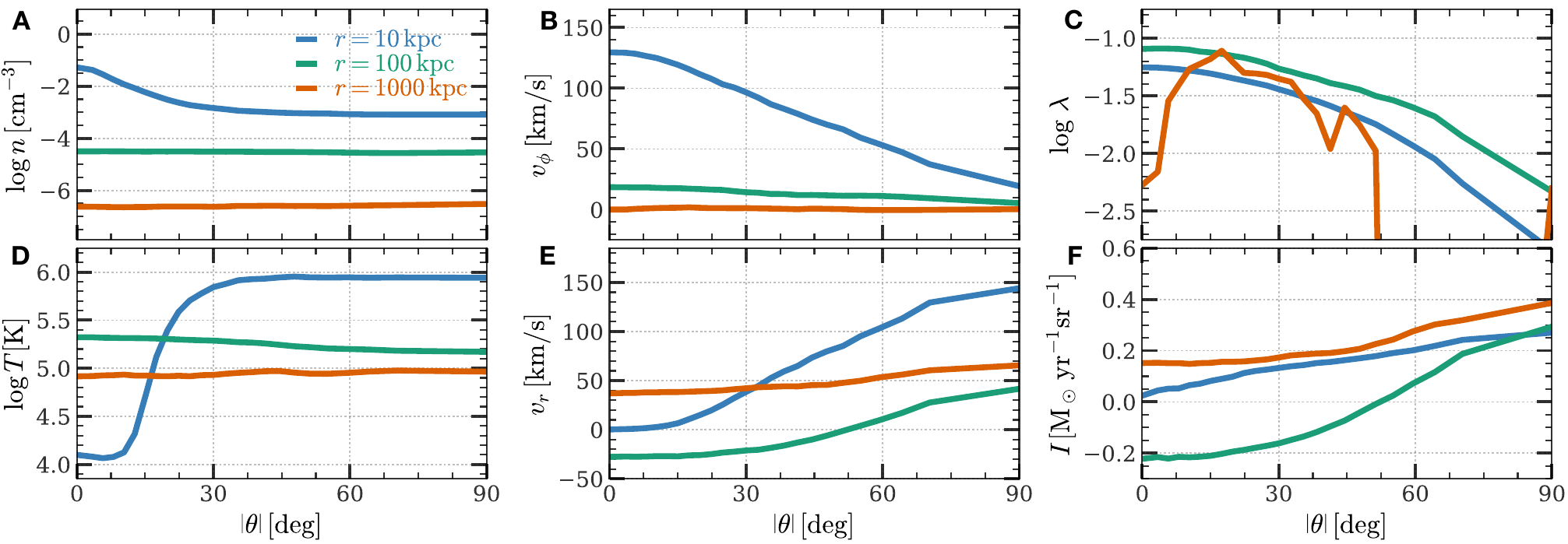}
    \caption{
    {\figem Angular profiles of gas properties.}
    This figure complements Figure~\ref{fig:mass10z0} by displaying
    how gas properties vary with (absolute value of) inclination angle relative to the disk plane 
    ($\left|\theta\right|$) at three representative galactocentric distances,
    $r = 10$, $100$ and $1000\Kpc$, which are 
    in the regimes of ISM, CGM and IGM, respectively.
    Azimuthal angle ($\phi$) is integrated out.
    }
   \label{fig:mass10z0-angular}
\end{figure*}

We applied the stacking method to construct 3D composite fields of 
gas properties (see Figure~\ref{fig:streamline} for the velocity field), and we further reduced them to
2D maps and 1D radial and angular profiles (Figure~\ref{fig:mass10z0} 
and Figure~\ref{fig:mass10z0-angular}). 
All fields exhibit pronounced radial gradients and angular anisotropy, 
with distinct features in different radial regimes (ISM, CGM and IGM) 
and angular zones (disk plane, outflow cone and recycling zone).

Gas density ($n$) and temperature ($T$) exhibit compatible 
patterns (Figure~\ref{fig:mass10z0}A,B). The density increases 
monotonically towards smaller radii, 
following an approximately isothermal profile ($n \propto r^{-2}$) 
through the regimes of IGM and CGM (Figure~\ref{fig:mass10z0}G). 
At $r \approx R_{\rm CGM}$, 
temperature begins to rise, indicating shock heating as gas flows into the local gravitational 
potential well. Within the ISM ($r \lesssim R_{\rm ISM}$), stalled azimuthal mixing creates 
prominent density anisotropy between the disk plane ($\theta = 0$) and outflow cone ($\left|\theta\right| = 90^\circ$). 
The outflow-cone density profile bends downward to be shallower than
the isothermal trend, while the disk-plane profile exhibits a sharp 
enhancement due to pile-up of infalling gas. 
The disk-plane density plateaus at $n \approx 0.1\,{\rm cm}^{-3}$ for $r \lesssim R_{\rm ISM}$, 
consistent with observed star-forming disks. 
The strong density concentration toward 
the disk plane at $r = 10\Kpc$ (Figure~\ref{fig:mass10z0-angular}A) 
confirms the geometric thinness of simulated disks, 
validating our alignment scheme used in stacking.
Temperature in the CGM remains nearly isothermal at $T \sim 10^5\,{\rm K}$, 
slightly below the halo virial temperature (e.g. Fig. 2 of ref.~\cite{moTwophaseModelGalaxy2024}). 
The disk plane is dominated by cooled gas at $T \approx 10^4\,{\rm K}$ set by the cooling 
barrier at around this temperature \cite{smithMetalCoolingSimulations2008}.
Inward of the disk, temperature increases mildly due to its density dependence
in the effective equation of state (eEOS) applied to star-forming gas
by IllustrisTNG \cite{springelCosmologicalSmoothedParticle2003,
pillepichSimulatingGalaxyFormation2018}.

The velocity field reveals the fountain pattern of the baryon cycle underlying 
the ensemble. 
Rotation velocity ($v_\phi$) exhibits strong anisotropy within the ISM, with 
rotation dominating in the disk plane but negligible in the 
outflow cone (Figure~\ref{fig:mass10z0}C,H) -- a direct 
consequence of the preference of gas with ordered motion to 
settle into the disk.
The disk-plane rotation velocity increases rapidly from $\lesssim 10\kms$ 
at $R_{\rm CGM}$ to $\sim 100\kms$ at $r \lesssim R_{\rm ISM}$, 
demonstrating that substantial angular momentum is preserved 
during radial transport. Within the ISM, $v_\phi$
approaches a plateau at $\lesssim 150\kms$ for $r \approx 10\,{\rm kpc}$, 
reflecting ongoing angular-momentum redistribution. 
This redistribution was found to occur through gravitational and 
magnetohydrodynamical torques from stars, dark matter and gas, enabling continuous 
angular-momentum transfer outward and allowing inward gas flow \cite{trappGasInfallRadial2022}. 
In the IGM, $v_\phi$ becomes negligible, indicating that large-scale bulk motion 
decouples from galactic rotation. 

The radial velocity $v_r$ also exhibits strong anisotropy. 
Along the disk plane, gas flows inward at velocities reaching 
$\sim 20\kms$ in the CGM, decelerating to a few $\kms$ in the ISM due to 
the angular-momentum barrier and depletion by star formation and feedback-driven 
outflows. The decelerated inflow reflects a scenario of accretion
limited by angular momentum and gas depletion, in contrast to
the cooling-limited accretion in the outer CGM (see Appendix~\ref{app:ssec:fountain-in-lcdm}).
The $\kms$-level inflow speed was found to be necessary to explain the observed 
SFRs in galaxies with this stellar mass,
as demonstrated analytically with a steady-state disk model \cite{wangOriginExponentialStarforming2022} 
and verified in high-resolution simulations resolving transport of angular 
momentum \cite{trappGasInfallRadial2022,trappAngularMomentumTransfer2024}.
In contrast, the outflow cone shows an accelerating outflow within the inner ISM
($r \lesssim 10\Kpc$), which converges to a near-constant velocity of 
$v_r \approx 50 \kms$ throughout the outer ecosystem (Figure~\ref{fig:mass10z0}D,K). Such a constant velocity indicates 
a balance between gravity and pressure gradients, and momentum conservation 
during radial transport of outflowing gas. The collimated geometry of the 
outflow throughout the CGM (Figure~\ref{fig:streamline}D,E;
\S\ref{ssec:outflow-geometry})
is a natural outcome of such a propagation with low resistance.
The velocity is transonic 
($\mathcal{M} = v_r/c_{\rm s} \approx 1$), producing only mild shock heating 
that is consistent with the smooth temperature profile (Figure~\ref{fig:mass10z0}J). 
The smooth density profile in the outflow cone, without a sharp drop, further 
demonstrates that stellar and thermal-mode AGN feedback primarily regulate star formation through 
reduced conversion efficiency rather than ejective quenching at this
stellar mass.  Beyond $R_{\rm CGM} \approx 300\Kpc$ 
(a few times $R_{\rm h}$), the Hubble expansion dominates over outflow velocity, 
limiting the impact of feedback at large scales. 

The spin parameter exhibits uniform values, with a factor-of-few variation, 
throughout the CGM regime near the 
disk plane ($\left|\theta\right| < 30^\circ$; Figure~\ref{fig:mass10z0}E, 
Figure~\ref{fig:mass10z0-angular}C), again indicating substantial 
preservation of angular momentum during radial transport of gas. 
Ref.~\cite{danovichFourPhasesAngularmomentum2015} attributed this conservation 
to negligible torques, though the underlying reasons remain incompletely understood. 
Candidate explanations include the inefficiency of dynamical friction from
low-density media on inflowing gas, and the alignment of gas 
with dark matter that minimizes gravitational torques. Further investigation, 
following the torque analysis of ref.~\cite{danovichFourPhasesAngularmomentum2015}, 
could separate these contributing factors.
At the ISM-CGM interface ($r \approx R_{\rm ISM}$), the spin parameter along the 
disk plane peaks at $\lambda \gtrsim 0.1$ 
(Figure~\ref{fig:mass10z0}I), signifying violent 
angular-momentum redistribution as infalling gas encounters the existing 
ISM. The spin exhibits marked angular anisotropy, with suppressed values in 
the outflow cone across the entire ecosystem (Figure~\ref{fig:mass10z0}I;
Figure~\ref{fig:mass10z0-angular}C). 
The spin-induced anisotropy of gas distribution starts to develop before gas 
reaches the ISM, as suggested by idealized simulations
\cite{hobbsFeedingSupermassiveBlack2011}. 
In contrast, $n$, $T$, and $v_\phi$ show weaker angular dependence in the 
CGM, indicating that mixing among gas elements remains incomplete 
(see Appendix~\ref{app:ssec:discussion-am} for more discussion)
and a spherical approximation is adequate to describe the hot accretion flows 
in this regime.

The mass flux $I$ mirrors the anisotropic pattern of radial velocity $v_r$, 
yet provides complementary insight. Although $v_r$ reaches 
only a few $\kms$ in the ISM along the disk plane, the mass flux there remains only 
about a factor of $2$ smaller than in the CGM (Figure~\ref{fig:mass10z0}L), 
reflecting the substantially higher gas density. Inward of 
$r \lesssim 10\Kpc$ along the disk plane, the mass 
flux declines as gas is consumed by star formation, expelled via feedback, 
or accumulated in the disk. Combined with velocity streamlines 
(Figure~\ref{fig:streamline}F), this pattern supports a scenario of {\bf\em coplanar inflow},
which was found to explain observed 
metallicity gradients in disk galaxies \cite{lyuDominantRoleCoplanar2025}. 
The flux reaching the nuclear region ($r \lesssim 1\Kpc$) remains small, 
implying slow growth of inner bulges and central SMBHs -- a consequence of 
our sample selection.
Beyond $r \gtrsim R_{\rm h}$, the outer CGM exhibits declining mass flux, 
indicating that the halos at $z \approx 0$ have transitioned to a slow-accretion 
phase \cite{zhaoGrowthStructureDark2003,moTwophaseModelGalaxy2024}. 
Consequently, most gas accretion of the central galaxies relies on 
the pre-existing CGM or that returned from the ISM.

\subsection{\label{app:ssec:fountain-in-lcdm}Build-up of baryon fountain in the $\Lambda$CDM paradigm}

The fountain pattern revealed in the stacked fields can be 
understood in the context of the $\Lambda$CDM structure-formation paradigm
\cite{moGalaxyFormationEvolution2010}. 
At $r \approx R_{\rm h}$ (dashed circles in Figure~\ref{fig:streamline}), 
rotational motion is already apparent from the non-zero impact parameters of the streamlines 
(Figure~\ref{fig:streamline}B).
Given a typical gas spin parameter of $\lambda \sim 0.1$ at this radius
(Figure~\ref{fig:mass10z0}I; see also Figure~1 of 
ref.~\cite{danovichFourPhasesAngularmomentum2015}), the azimuthal velocity at 
the halo boundary is roughly
\begin{equation}
    v_\phi \sim \sqrt{2} \lambda V_{\rm h} \sim 0.14 V_{\rm h} \sim 14\kms \,,
\end{equation}
consistent with the color scale of the streamlines.
The radial motion is driven by the gravitational attraction of the local 
over-density acting on the radiatively cooled gas.
Approximating the inflow as a steady-state, cooling-limited flow, which is 
an appropriate description for our ensemble at $r\sim R_{\rm h}$
where the CGM is largely virialized \cite{goldnerAccretionDrivenTurbulenceCircumgalactic2025},
the radial velocity can be estimated as
\begin{multline}
    v_r \approx -\frac{r}{t_{\rm cool}} \approx
    -12\kms \left( \frac{M_{\rm h}}{10^{12}\Msun} \right)^{-0.36}\left(\frac{\dot{M}}{1\msunperyr}\right)^{1/2} \\
    \times \left(\frac{\Lambda_{\rm cool}}{10^{-22}\,{\rm erg\,cm^3\,s^{-1}}}\right)^{1/2}
	\left(\frac{r}{150\Kpc}\right)^{-0.4}
    \,,\label{eq:v-r-estimate}
\end{multline}
where we adopted an analytic expression for the cooling timescale (eq.~30 of 
ref.~\cite{sternCoolingFlowSolutions2019});
$\dot{M}$ is the mass inflow rate (with a typical value of $\sim 1\msunperyr$; 
see Figure~\ref{fig:mass10z0}L);
and $\Lambda_{\rm cool}$ is the cooling function evaluated near 
$10^5\Kelvin$ \cite{smithMetalCoolingSimulations2008}.
The inflow velocity so obtained agrees with the streamline directions in 
Figure~\ref{fig:streamline}B.
The characteristic velocity dispersion of the gas, whether in thermal or 
turbulent form,
is of order $\sigma_v \sim (GM_{\rm h}/R_{\rm h})^{1/2}\sim V_{\rm h}$, which is 
much larger than both $v_\phi$ and $|v_r|$. 
Hence, gas kinematics near the halo boundary is dominated 
by random motion. A simple estimate of the gas density in this regime gives
\begin{equation}
    n \lesssim \Delta_{\rm v}\,\rho_{\rm c}\,\Omega_{\rm B}
    \approx 4\times 10^{-5}\perccm\,,
\end{equation}
well below the star-formation threshold ($\sim 0.1\perccm$) adopted in 
TNG \cite{pillepichSimulatingGalaxyFormation2018}. 
Nevertheless, at these densities the cooling timescale becomes comparable to 
or shorter than the free-fall timescale ($t_{\rm cool} \lesssim t_{\rm ff}$; see e.g. 
Figure~2 of \cite{moTwophaseModelGalaxy2024}), allowing shock-heated 
gas to cool and flow inward. Thus, halo formation establishes a gravity-driven, 
cooling-limited, random-motion-dominated regime in the outer CGM that acts as 
a reservoir supplying gas to the inner ecosystem.

At larger radii, cosmic expansion overwhelms the gravity-driven inflow. 
With $H_0 \approx 70 \kms{\rm Mpc}^{-1}$ and typical inflow speed 
of $\sim 10 \kms$ (eq.~\ref{eq:v-r-estimate}), 
$v_r$ is expected to reverse at $r\sim$ a few hundred kpc. This reversal 
is visible in the face-on slice of velocity field (Figure~\ref{fig:streamline}A) 
as a ring-shaped ridge, and the scale of this reversal is commonly referred to as 
the turnaround scale \cite{pavlidouWhereWorldStands2014,leeTurningCosmicWeb2016}. 
Outside this ridge, matter is unlikely to be accreted directly into the central 
galaxy; nevertheless, the intergalactic environment can still influence galaxy 
evolution gravitationally, e.g. by torquing dark matter \cite{wangInternalPropertiesEnvironments2011} 
and gas streams \cite{danovichFourPhasesAngularmomentum2015}, 
or radiatively \cite{wiseFormationMassiveBlack2019,latifUVRegulatedStar2019}, 
and can receive responses from galaxies through feedback-driven outflows. 
This partial decoupling helps explain subtle correlations between galaxies 
and their large-scale environments, such as two-halo 
galactic conformity \cite{kauffmannReexaminationGalacticConformity2013,
hearinHaloMassGalactic2015,wangDissectTwohaloGalactic2023} 
and environmental dependence of galactic structures
\cite{liDependenceClusteringGalaxy2006,zhangUnexpectedClusteringPattern2025}.

At smaller radii, the importance of rotation rises. If a substantial fraction 
of angular momentum is retained in the cooling flow (see Appendix~\ref{app:ssec:discussion-am}), 
then $v_\phi\propto r^{-1}$ while the 
inflow speed follows approximately $|v_r|\propto r^{-0.4}$ (eq.~\ref{eq:v-r-estimate}). 
The difference in the radial dependence of $v_r$ and $v_\phi$ produces the radial change of 
streamline direction seen in the face-on view (Figure~\ref{fig:streamline}B). 
When the gas contracts by roughly $1/\lambda\sim 10$, rotational support becomes 
comparable to that provided by random motion. Azimuthal mixing stalls, and
the gas flattens into a disk. The resulting rise in density (to $\sim 0.1\perccm$) makes the gas 
locally self-gravitating, triggers gravitational instability, and enables star 
formation. Our definition of the ISM regime based on the $v_\phi$ profile therefore 
naturally captures the physical meaning of ``interstellar''. Baryons inside 
this regime are primarily locked into collisionless stars and collisional, 
interstellar gas that interacts frequently with stars via gravity and feedback. 
Because azimuthal mixing is suppressed, radial inflow along the disk is reduced 
and bulk motion within the ISM becomes dominated by rotation rather than by 
radial migration or turbulent/thermal motion, as evident in Figure~\ref{fig:streamline} 
on scales of a few tens of kpc. In summary, the ISM regime is a 
self-gravity-driven, rotation-dominated, star-forming regime distinct 
from the outer regimes.

Along the disk plane, the CGM shows net inward motion at 
$\lesssim 30\kms$, feeding the disk via anisotropic mixing. Perpendicular 
to the disk, a bipolar outflow cone is apparent, with 
outflow velocities sustained at $\approx 50\kms$ out to CGM and IGM scales. 
As a crude estimate, the radial reach of supernova (SN)-driven outflow
in one dynamical timescale of the host halo can be 
approximated by the blast-wave solution \cite{chenTwophaseModelGalaxy2025a}:
\begin{equation}
    \frac{R_{\rm sh}}{R_{\rm h}} \approx 2.76 \chi_{\rm sh}^{1/5} 
    \left(\frac{M_{\rm h}}{10^{12}\Msun}\right)^{-0.105}
    \,,\label{eq:R-sh-over-R-h}
\end{equation}
where $R_{\rm sh}$ is the shock radius, and the order-unity parameter $\chi_{\rm sh}$ encapsulates factors 
such as energy injection efficiency of SNe, star-formation efficiency and local 
gas over-density. This scaling implies $R_{\rm sh}\propto R_{\rm h}$ to first order,
with very weak mass dependence. 
For our fiducial ensemble, $R_{\rm sh}$ so estimated is $ \sim 400\Kpc$,
large enough to encompass the volume enclosed by $R_{\rm CGM}$ computed 
for this ensemble.
With additional energy provided by the thermal-mode AGN, the outflow 
may be sustained further to larger radii. 

At intermediate inclinations ($|\theta|\sim 45^\circ$), outflowing gas turns around 
in the CGM regime, recycling material back to the galaxies. In IllustrisTNG, stellar winds and 
thermal-mode AGN feedback are injected isotropically on small 
scales \cite{weinbergerSimulatingGalaxyFormation2017,pillepichSimulatingGalaxyFormation2018}.
The bipolar geometry of the large-scale outflow therefore arises from hydrodynamic 
coupling with the anisotropic ambient medium. The dense disk plane confines 
the outflow, while the polar directions provide paths with less resistance,
thus shaping a cone-like outflow geometry.
Such a coupling-induced anisotropy pattern appears to be generic: simulations 
with additional modeling of multiphase ISM were also found to produce bipolar outflow 
around disks but not around galaxies with near-spherical gas 
distribution \cite{sivasankaranAGNFeedbackIsolated2025}. 
The detailed structure within this generic pattern may vary in individual
ecosystems, and may depend on the sub-grid prescriptions implemented in 
simulations. For example, a high-energy, bipolar AGN jet at injection can 
produce a collimated outflow even without 
confinement by the galactic disk, as implied by jets observed around 
galaxies \cite{nulsenClusterScaleAGNOutburst2005,
fabianObservationalEvidenceActive2012,oeiBlackHoleJets2024}; 
the degree of collimation and the opening angle of the outflow cone
can vary with jet parameters and conditions of ambient medium, 
as suggested by hydrodynamic 
simulations \cite{suSelfregulationBlackHole2023,duttaDissipationAGNJets2024,
suSelfregulationHighredshiftBlack2025}.

\subsection{\label{app:ssec:discussion-am}The acquisition and redistribution of angular momentum}

\begin{figure*} \centering
    \includegraphics[width=0.5\textwidth]{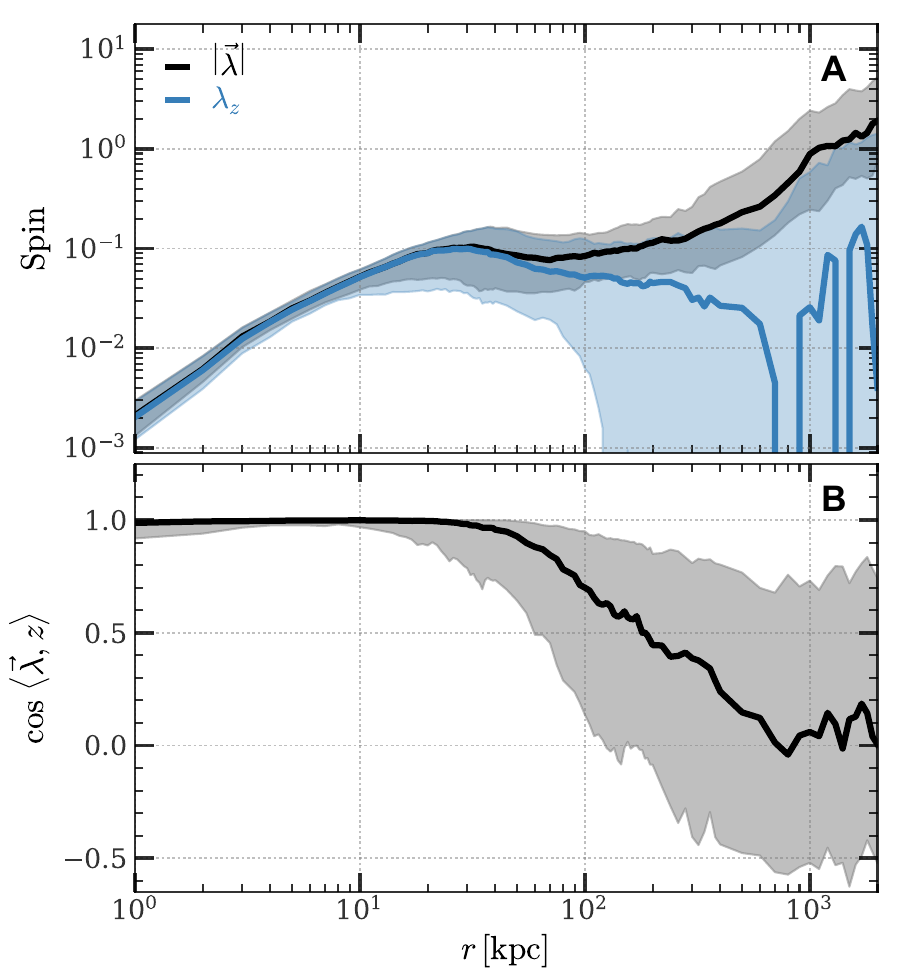}
    \caption{
    {\figem Profiles of spin amplitude and alignment.}
    {\figem (A)} Magnitude of spin vector ($|\vec{\lambda}|$), and its component 
    along the disk normal ($\lambda_z$).
    {\figem (B)} Alignment of the spin vector with the disk normal, quantified 
    by the cosine of their angle.
    Profiles are spherically averaged (i.e. evaluated in spherical shells centered
    on each galaxy), and are shown as functions of galactocentric distance ($r$).
    Solid curves show the median among galaxies in the fiducial ensemble,
    with shading indicating the $16^{\rm th}$--$84^{\rm th}$ percentile range.
    Both panels reveal imperfect angular-momentum alignment in the CGM 
    regime across radial shells, indicating that the mixing process has 
    not yet been completed to homogenize spin distribution 
    before gas reaches the ISM.
    }
    \label{fig:spinalignment}
\end{figure*}

The plateau of the spin profile in the CGM regime reflects the 
preservation of angular momentum during inflow. 
This can be understood by noting that the CGM spin parameter, 
$\lambda_0 \approx 0.1$, is similar to that of recently accreted gas, 
and such spin evolves slowly at $z \lesssim 2$ (see e.g. Figures~2 and 5 of 
\cite{stewartAngularMomentumAcquisition2013}).
To exclude the alternative explanation that this plateau results from a mixing process that homogenizes 
the angular momenta of gas elements, we evaluate the average spin 
vector ($\vec{\lambda}$) of gas elements in spherical shells at different 
galactocentric distances ($r$). We show the magnitude of the 
spin vector ($|\vec{\lambda}|$) as a function of $r$ 
in Figure~\ref{fig:spinalignment}(A), 
and compare it with the spin component along the disk normal direction 
($\lambda_z$) for our fiducial ensemble.
We further quantify angular-momentum alignment across radii by computing 
the angle between $\vec{\lambda}$ and the disk normal, and 
showing it in Figure~\ref{fig:spinalignment}(B).
If the mixing process dominates, we expect gas shells at different radii 
to have aligned angular-momentum vectors.
However, the alignment between $\vec{\lambda}$ and the disk normal weakens 
with increasing $r$ in the CGM regime, reaching a misalignment of  
$\sim 60^\circ$ at $r = R_{\rm h}$.
This indicates that mixing remains far from complete in the outer CGM, 
rather than homogenizing the spin distribution.

At the ISM-CGM interface, the angular-momentum barrier decelerates 
infalling gas, causing it to accumulate and form a dense reservoir 
(evident in the sharp rises of $n$ and $v_\phi$, and decline of $v_r$
at $r \sim R_{\rm ISM}$ in Figure~\ref{fig:mass10z0}).
This provides sufficient time for the mixing process to complete, producing
perfectly aligned angular momenta (Figure~\ref{fig:spinalignment}B), 
a flattened, condensed disk, and a rapid drop in temperature (Figure~\ref{fig:mass10z0}).
High-speed collisions between inflowing and pre-existing gas, 
supplemented by energy injection from feedback \cite{krumholzTurbulenceInterstellarMedium2016,bacchiniEvidenceSupernovaFeedback2020},
generate substantial turbulence in this regime \cite{vollmerQuenchingStarFormation2013,goldnerAccretionDrivenTurbulenceCircumgalactic2025}. 
This turbulence continuously produces low-angular-momentum gas elements 
that sink inward \cite{vollmerQuenchingStarFormation2013}, making the inner CGM 
simultaneously a source of matter for the ISM and a sink of angular momentum 
that stores rotational energy redistributed by sinking gas.
The peak of the spin profile at $r \approx R_{\rm ISM}$ (Figure~\ref{fig:mass10z0}I) 
confirms the role of inner CGM as an angular-momentum reservoir.
Future simulation studies could test these mechanisms by tracing individual gas elements 
and computing their angular-momentum evolution throughout their trajectories
in galaxy ecosystems. This remains technically challenging in moving-mesh codes due 
to cell merging and splitting, but is more feasible in smoothed-particle-hydrodynamics (SPH)
simulations \cite{springelCosmologicalSmoothedParticle2002,springelCosmologicalSimulationCode2005,schayeEAGLEProjectSimulating2015,schayeCOLIBREProjectCosmological2026}.
Notably, the misaligned spin distribution in the CGM and the violent 
angular-momentum redistribution at the ISM-CGM interface lead to a prediction: 
if a large amount of infalling gas with significantly misaligned spin enters the ISM, 
the disk could be substantially warped or destroyed, potentially triggering 
bulge formation and AGN activity \cite{sankarHotAccretionSpiral2025}.
Recent observations with integral-field-unit (IFU) spectroscopy for misaligned galaxies 
have provided evidence for such processes \cite{zhouMisalignedGasAcquisition2024,
zhouMisalignedExternalGas2025,baoDifferentInfluenceGas2024,zhangImpactExternalGas2025}.

\subsection{\label{app:ssec:model}The parametric model for baryon cycle}

We developed a parametric model to quantitatively describe the baryon cycle in 
ensembles of galaxies. While the large spatial scale of the baryon cycle and diverse 
physical processes involved necessitate a complex model, the regular patterns 
and clear divisions revealed by our stacking method allow for a composable approach. 
Each component, corresponding to a specific regime, can be separately 
constructed and calibrated. 
Here, we focus on modeling the velocity field around disk galaxies, 
though the methodology is generalizable to other properties and ensembles. 
Our priority is physical interpretability and extensibility rather than mathematical 
simplicity. The model was fitted to the stacked velocity field of the fiducial 
ensemble, and the best-fit parameters and their physical implications are 
discussed below.

The rotational motion in the disk plane is captured by parameterizing 
the spin profile, $\lambda$ (see Figure~\ref{fig:mass10z0}I), as
\begin{equation}
    \lambda(\tilde{r}) = 
    \begin{cases}
        \left(\lambda_{\rm p}-\lambda_0\right)\mathcal{S}(\tilde{r}; \tilde{r}_\lambda, \sigma_\lambda, 
            \alpha_\lambda) + \lambda_0 \,, 
        & \text{if } \tilde{r} \geqslant \tilde{r}_\lambda \,; \\
        \lambda_{\rm p} \mathcal{P}(\tilde{r}; \tilde{r}_\lambda, \tilde{r}_\lambda, \beta_\lambda) \,,
        & \text{otherwise} \,,
    \end{cases}
    \label{eq:model-lambda-disk}
\end{equation}
with the rotational velocity given by $v_\phi(\tilde{r}) = \sqrt{2} R_{\rm h} V_{\rm h}\lambda(\tilde{r}) / r$. 
Here, $\tilde{r} \equiv r/R_{\rm h}$ is the dimensionless galactocentric 
radius. 
The modified exponential function is defined as
\begin{equation}
 \mathcal{S}(x; x_0, \sigma, \alpha) \equiv 
 \exp\left[-\left(\frac{x - x_0}{\sigma}\right)^{\alpha+1}\right]\,,
\end{equation}
where $x_0$ is the center, $\sigma$ the width, and $\alpha$ the sharpness parameter. 
The modified power-law function is
\begin{equation}
 \mathcal{P}(x; x_0, \Delta x, \alpha) \equiv \frac{1}{\Delta x^{\alpha+1}} 
 \left[ \Delta x^{\alpha+1} - (x_0 - x)^{\alpha+1} \right]\,,
\end{equation}
where $x_0$ is the anchor, $\Delta x$ the width, and $\alpha$ the sharpness. 
The two pieces of $\lambda(\tilde{r})$ capture the large-scale plateau and 
the inner rise of the spin profile, respectively, with a smooth bump of 
$\lambda = \lambda_{\rm p}$ in between at $\tilde{r}_\lambda \sim R_{\rm ISM}/R_{\rm h}$.
The best-fit parameters are: $\lambda_{\rm p} = 0.120$ (peak spin of the bump), 
$\lambda_0 = 0.082$ (spin plateau in the outer CGM), 
$\tilde{r}_\lambda = 0.191$,
$\sigma_\lambda = 0.144$ (indicating angular momentum redistribution at 
$\sim 0.2 R_{\rm h}$ over a thin layer of thickness $\sim 0.1 R_{\rm h}$), 
$\alpha_\lambda = 1.151$ (indicating a Gaussian-like profile in the 
redistribution layer), and
$\beta_\lambda = 0.529$ (indicating a slight convexity of the inner 
rise part).

The inflow along the disk plane ($\theta = 0$; Figure~\ref{fig:mass10z0}K) is parameterized as
\begin{multline}
    \tilde{v}_r(\tilde{r}) = 
    - \tilde{v}_{\rm in} \tilde{r}^{-\alpha_{\rm in}} 
    \mathcal{P}^\text{+}(\tilde{r};\tilde{r}_{\rm in}; \Delta \tilde{r}_{\rm in},\beta_{\rm in}) \\
    - \tilde{v}_{\rm d} \left(\frac{\tilde{r}}{\tilde{r}_{\rm d}}\right)^{\alpha_{\rm d}}
     \mathcal{S}(\tilde{r}; 0, \tilde{r}_{\rm d}, \beta_{\rm d}) 
    + \tilde{v}_{\rm H}(\tilde{r}) 
    \,,\label{eq:model-v-r-disk}
\end{multline}
where $\tilde{v}_r \equiv v_r / V_{\rm h}$; 
$\mathcal{P}^\text{+}(\tilde{r}) \equiv \max\left[\mathcal{P}(\tilde{r}),0\right]$
at $\tilde{r} \leqslant \tilde{r}_{\rm in}$ and $1$ otherwise;
$\tilde{v}_{\rm H}(\tilde{r}) \equiv H(z) t_{\rm dyn}(z) \tilde{r}$ is the Hubble 
flow; $t_{\rm dyn}(z) \equiv R_{\rm h}/V_{\rm h}$ is the dynamical timescale of 
the halo. The first term describes cooling-limited inflow in the outer CGM 
(motivated by eq.~\ref{eq:v-r-estimate}), with the cutoff function $\mathcal{P}^\text{+}$ 
representing angular-momentum deceleration in the inner CGM. The second term 
models the transition to angular-momentum-limited inflow within the disk. The 
third term accounts for Hubble expansion, which dominates in the IGM. The best-fit 
parameters are: 
$\tilde{v}_{\rm in}=0.277$ (indicating a velocity $\lesssim 0.3V_{\rm h}$ at the 
halo boundary), 
$\alpha_{\rm in}=0.510$ (close to the analytic value of $0.4$; see eq.~\ref{eq:v-r-estimate}), 
$\tilde{r}_{\rm in}=0.817$ (the radius at which the angular-momentum barrier starts 
to decelerate the inflow), 
$\Delta \tilde{r}_{\rm in} = 0.626$ (the barrier is fully established 
over $\sim 0.6 R_{\rm h}$ and halts the inflow at $\sim 0.2 R_{\rm h}$,
consistent with $\tilde{r}_{\lambda}$ in eq.~\ref{eq:model-lambda-disk}), 
$\beta_{\rm in}=1.584$, 
$\tilde{v}_{\rm d}=0.113$ (indicating a slow inflow in the disk), 
$\tilde{r}_{\rm d}=0.020$ (consistent with typical disk size; see e.g. \cite{somervilleRelationshipGalaxyDark2018,panRoleRegulatingSize2021}), 
$\alpha_{\rm d}=0.251$, 
and $\beta_{\rm d}= 0.001$ (indicating a nearly exponential inflow profile).

Using the best-fit $\lambda$ and $\tilde{v}_r$ profiles, we reconstructed the face-on 
velocity field in the disk plane (Figure~\ref{fig:model}B). The model reproduces the 
main features of the stacked field (Figure~\ref{fig:model}C): large-scale Hubble 
flow in the IGM, a ridge of divergent flows at $R_{\rm CGM}$, spiraling-in flow 
in the CGM, and a sharp rise in rotation within the ISM.

The radial velocity along the outflow-cone center ($\left|\theta\right| = 90^\circ$; Figure~\ref{fig:mass10z0}K) 
is parameterized as
\begin{equation}
    \tilde{v}_r(\tilde{r}) = 
    \left\{
    \begin{array}{lr}
        \left(\tilde{v}_{\rm out,p} - \tilde{v}_{\rm out,0}\right) 
        \mathcal{S}(\tilde{r}; \tilde{r}_{\rm out,p}, \sigma_{\rm out}, \alpha_{\rm out})
        \\
        \ \ \ \ \ \ \ \ \ \ \ \ \ \ \ \ \ \ \ \ \ \ \ \ \ 
        + \tilde{v}_{\rm out,0} \,,\text{if } \tilde{r} \geqslant \tilde{r}_{\rm out,p}\\
        \tilde{v}_{\rm out,p} 
        \mathcal{S}(\tilde{r}; \tilde{r}_{\rm out,p}, \delta_{\rm out}, \beta_{\rm out})
        \,,\text{otherwise}
    \end{array}
    \right\} + \tilde{v}_{\rm H}(\tilde{r})
    \,,
    \label{eq:model-v-r-cone}
\end{equation}
where the two pieces capture the stabilized outflow velocity
of $\tilde{v}_{\rm out,0}$ at large scales and the 
acceleration of outflow in the inner ecosystem, 
respectively, with a transitional bump of $\tilde{v}_{\rm out,p}$
in between at $\tilde{r}_{\rm out,p}$.
The best-fit parameters are: 
$\tilde{r}_{\rm out,p}=0.026$, 
$\tilde{v}_{\rm out,p}=1.727$
(indicating that the outflow accelerates to a peak velocity of 
$\sim 1.7 V_{\rm h}$ in the inner ISM of $r \sim 0.03 R_{\rm h}$,
which is comparable to the disk size $\tilde{r}_{\rm d}$ obtained in 
eq.~\ref{eq:model-v-r-disk}),
$\delta_{\rm out}=0.010$, 
$\beta_{\rm out}=0.079$ (characterizing the shape of the profile), 
$\tilde{v}_{\rm out,0} =0.342$,
$\alpha_{\rm out}=-0.003$, 
$\sigma_{\rm out}=0.130$ (indicating that the outflow converges to a 
near-constant velocity of $\sim 0.3 V_{\rm h}$ after passing a narrow radial 
range of $\gtrsim 0.1 R_{\rm h}$). The angular dependence of $v_r$ is modeled as 
a mixture of inflow and outflow:
\begin{align}
    \tilde{v}_r(\tilde{r}, \theta) = 
    &\left[\tilde{v}_r(\tilde{r},0) - \tilde{v}_{\rm H}(\tilde{r})\right]
    \mathcal{S}(\left|\theta\right|; 0, \sigma_{r\theta}, \alpha_{r\theta})
    \nonumber\\
    &\,\,\,\,\,
    +\left[\tilde{v}_r(\tilde{r},\frac{\pi}{2}) - \tilde{v}_{\rm H}(\tilde{r})\right]
    \mathcal{S}(\left|\theta\right|; \frac{\pi}{2}, -\delta_{r\theta}, \beta_{r\theta})
    \nonumber\\
    &\,\,\,\,\,
    +\tilde{v}_{\rm H}(\tilde{r})
    \,,
\end{align}
where $\tilde{v}_r(\tilde{r},0)$ and $\tilde{v}_r(\tilde{r},\pi/2)$ are the disk-plane 
and outflow-cone velocities from eqs.~\ref{eq:model-v-r-disk} and 
\ref{eq:model-v-r-cone}. The best-fit parameters are: $\alpha_{r\theta}=5.000$, 
$\sigma_{r\theta}=1.186$, $\beta_{r\theta}=1.238$, $\delta_{r\theta}=0.769$, 
describing the angular geometry of radial flow (see Figure~\ref{fig:mass10z0}D). 
A more intuitive representation of the outflow geometry is the opening angle 
of the outflow cone, as shown in Figure~\ref{fig:outflow-geometry}. 

The angular velocity $v_\theta$ is modeled as two components of opposite sign 
to capture the cycling pattern:
\begin{equation}
    \tilde{v}_\theta(\tilde{r}, \theta) = 
    \sum_{i = 1,2}
    \tilde{v}_{\theta}^{(i)} (\tilde{r}) f_\theta^{(i)}(\theta)
    \,.
\end{equation}
The radial dependence is
\begin{equation}
    \tilde{v}_{\theta}^{(i)}(\tilde{r}) = 
    \tilde{v}_{\theta r}^{(i)} \times \left\{
    \begin{array}{lr}
        \mathcal{S}(\tilde{r}; \tilde{r}_\theta^{(i)}, 
            \sigma_{\theta r}^{(i)}, \alpha_{\theta r}^{(i)})
        \,,&\text{if } \tilde{r} \geqslant \tilde{r}_\theta^{(i)}\\
        \mathcal{S}(\tilde{r}; \tilde{r}_\theta^{(i)}, 
            -\delta_{\theta r}^{(i)}, \beta_{\theta r}^{(i)})
        \,,&\text{otherwise}
    \end{array}
    \right\} + \tilde{v}_{\rm H}(\tilde{r})\,,
\end{equation}
with $\tilde{v}_{\theta r}^{(1)}< 0$ and $\tilde{v}_{\theta r}^{(2)} > 0$. 
The angular dependence is
\begin{equation}
    f_\theta^{(i)}(\theta)
    = A^{(i)} \text{sign}(\theta) \left|\theta\right|^{1+\alpha_{\theta\theta}^{(i)}}
        \left(\frac{\pi}{2}-\left|\theta\right|\right)^{1+\beta_{\theta\theta}^{(i)}}
    \,,
\end{equation}
where $A^{(i)}$ normalizes $f_\theta^{(i)}(\theta)$ to unity at 
its maximum for $\theta \in [-\pi/2,\pi/2]$, and $\text{sign}(\theta)$ is the sign 
function. The best-fit parameters are: 
$\tilde{v}_{\theta r}^{(1)}=-0.075$, 
$\tilde{r}_\theta^{(1)}=2.475$ (indicating a maximum $\left|v_\theta\right| \sim 0.1V_{\rm h}$ 
towards the disk plane is reached at large scales of $r \sim 2.5 R_{\rm h}$,
recycling the outflowing gas back to the disk),
$\sigma_{\theta r}^{(1)}=9.919$, 
$\delta_{\theta r}^{(1)}=1.216$, 
$\alpha_{\theta r}^{(1)}=0.545$, 
$\beta_{\theta r}^{(1)}=2.579$, 
$\tilde{v}_{\theta r}^{(2)}=0.364$, 
$\tilde{r}_\theta^{(2)}=0.065$ (indicating a maximum $v_\theta \sim 0.4 V_{\rm h}$ 
towards the outflow cone is reached at small scales of $r \sim 0.1 R_{\rm h}$,
refilling the outflow cone),
$\sigma_{\theta r}^{(2)}=0.123$, 
$\delta_{\theta r}^{(2)}=0.042$, 
$\alpha_{\theta r}^{(2)}=0.001$, 
$\beta_{\theta r}^{(2)}=4.493$, 
$\alpha_{\theta\theta}^{(1)} =0.169$, 
$\beta_{\theta\theta}^{(1)} =0.525$, 
$\alpha_{\theta\theta}^{(2)} =1.299$, 
$\beta_{\theta\theta}^{(2)} =0.051$.

Using the best-fit models for $\tilde{v}_r(\tilde{r},\theta)$ and 
$\tilde{v}_\theta(\tilde{r},\theta)$, we reconstructed the edge-on velocity field 
(Figure~\ref{fig:model}E). The model successfully recovers the butterfly-shaped 
pattern and the distinct zones of inflow, outflow and recycling.
The residuals between the model and simulation (Figure~\ref{fig:model}G), 
particularly within the ISM and the outflow cone, are more pronounced than 
in the face-on view, 
reflecting the rich substructure arising from the launch of outflow
and its subsequent interaction with the ambient medium. 
Asymmetries in the residuals relative to the disk plane are attributable to cosmic variance from the finite 
ensemble size. The ability of the model to reproduce the main features of 
the velocity field demonstrates that quantitative characterization of the 
baryon cycle is feasible when a regular pattern is revealed by stacking.

\subsection{\label{app:ssec:mass-dependence}Mass dependence of baryon cycle}

\begin{figure*}[t]
    \centering
    \includegraphics[width=\textwidth]{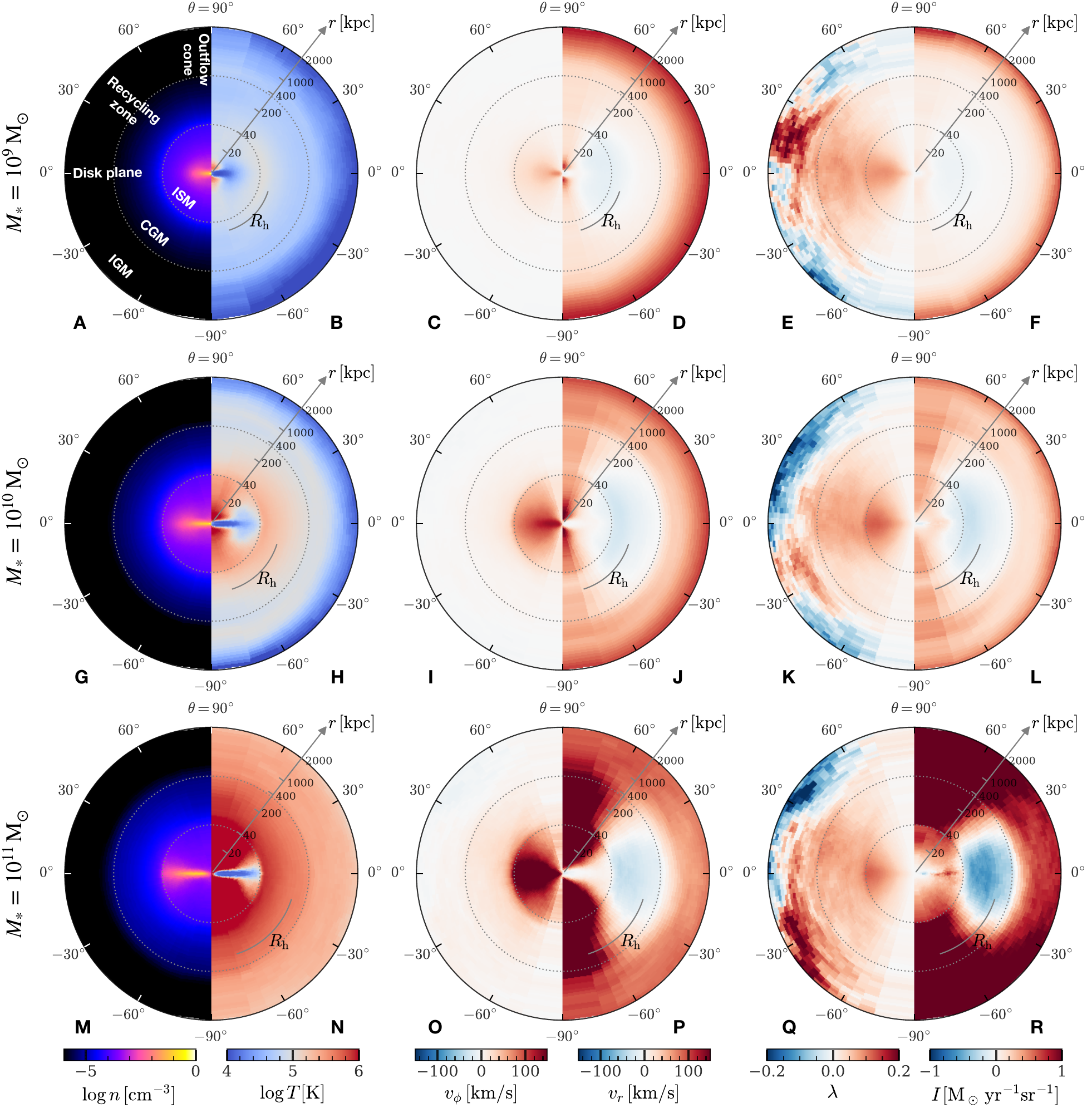}
    \caption{
    {\figem Stacked 2D maps of gas properties around ensembles of galaxies 
    with different stellar masses.} 
    This figure is analogous to the first row of Figure~\ref{fig:mass10z0}, 
    but comparing gas properties across three ensembles of star-forming
    galaxies at $z=0$ with 
    {\figem (A)}--{\figem (F)}, $M_* \approx 10^9$, 
    {\figem (G)}--{\figem (L)}, $10^{10}$, 
    and {\figem (M)}--{\figem (R)}, $10^{11} \Msun$. 
    The piecewise-linear scaling for the radial coordinate ($r$) 
    is consistent across all panels, with annotated regimes of ISM, CGM, 
    and IGM fixed to match those for the ensemble with 
    $M_* \approx 10^{10} \Msun$. 
    This figure illustrates that the overall configuration of the 
    baryon fountain is maintained across galaxies with different $M_*$, 
    while the physical extents of the relevant regimes co-expand 
    with the halo radius ($R_{\rm h}$) of host halos, 
    as indicated by the arcs. 
    See Figure~\ref{fig:mdependence-curves} for the 
    corresponding 1D profiles.
    }
   \label{fig:mdependence}
\end{figure*}

\begin{figure*}[t] 
	\centering
    \includegraphics[width=\textwidth]{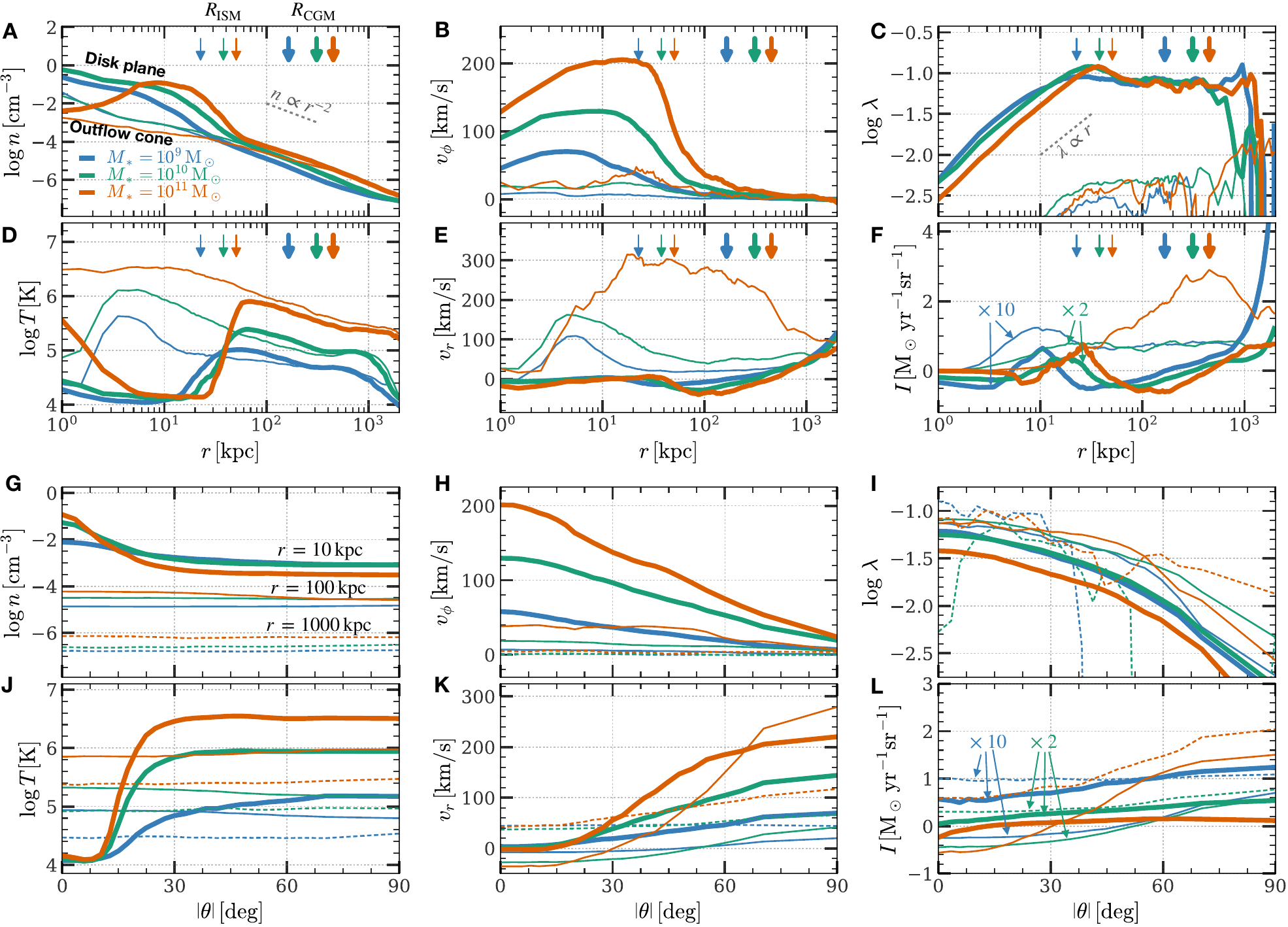}
    \caption{
    {\figem Stacked 1D profiles of gas properties around ensembles of galaxies 
    with different stellar masses.} 
    {\figem (A)}--{\figem (F)} Radial profiles along the disk plane 
    (thick curves) and outflow-cone center (thin curves).
    {\figem (G)}--{\figem (L)} Angular profiles at three representative radii.
    This figure is analogous to those shown in Figure~\ref{fig:mass10z0} and 
    Figure~\ref{fig:mass10z0-angular}, but for three ensembles of 
    star-forming galaxies with different $M_*$.
    Thin and thick arrows mark the ISM-CGM interface ($R_{\rm ISM}$) and CGM-IGM 
    interface ($R_{\rm CGM}$), respectively, for each ensemble.
    Profiles at $M_* \approx 10^9$ and $10^{10}\Msun$ in 
    {\figem (F)}, {\figem (L)} are multiplied by factors of $10$ and $2$, 
    respectively, to show the similarity of profiles among the ensembles.
    See Figure~\ref{fig:mdependence} for corresponding 2D maps.
    }
    \label{fig:mdependence-curves}
\end{figure*}

\begin{figure*}[t]
	\centering
    \includegraphics[width=\textwidth]{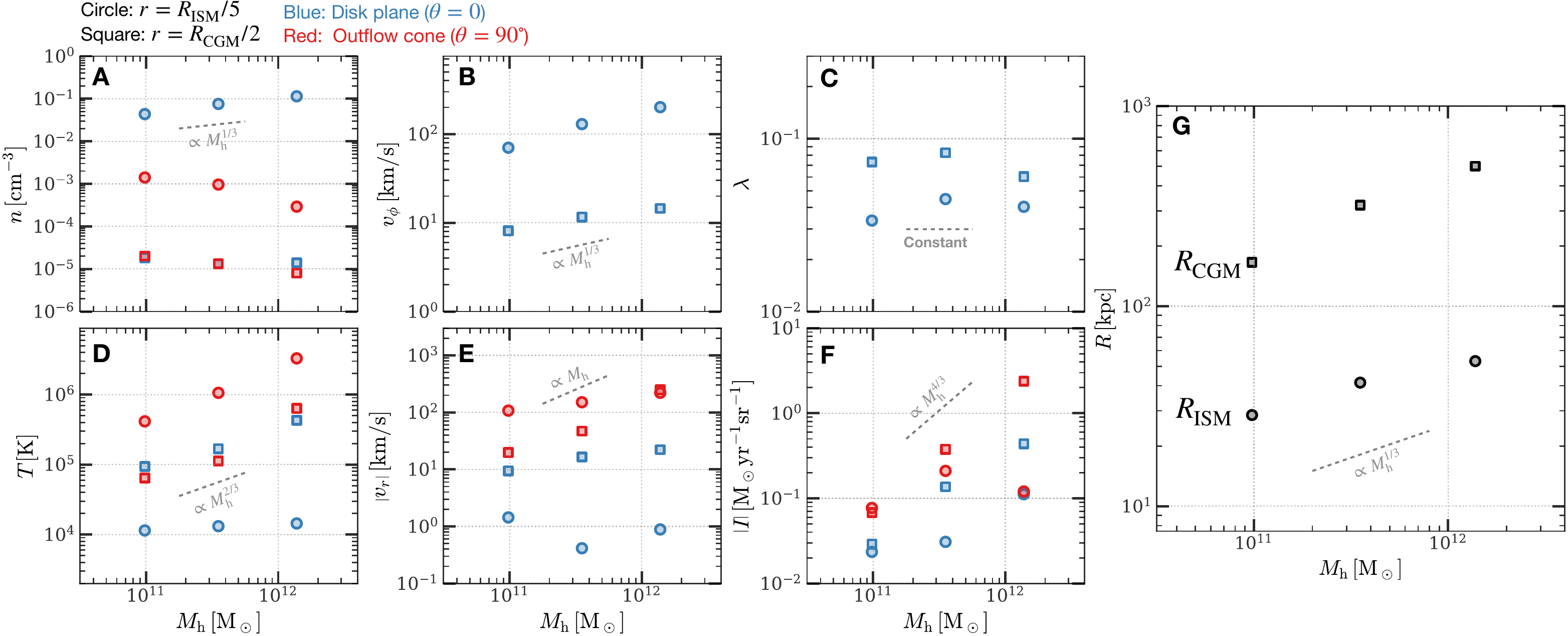}
    \caption{
    {\figem Role of halos in shaping galaxy ecosystems.}
    {\figem (A)}--{\figem (F)} Gas properties as functions of halo mass.
    Circles and squares show the values computed at $r = R_{\rm ISM}/5$
    and $R_{\rm CGM}/2$, respectively, along the disk plane (blue) 
    or outflow-cone center (red).
    {\figem (G)} Scales ($R_{\rm ISM}$ and $R_{\rm CGM}$) of the galaxy ecosystem 
    as functions of halo mass.
    The results here are obtained from the ensembles at $z=0$ with 
    different masses (IDs=1, 2, 3; see 
    Table~\ref{tab:subsamples}).
    For reference, each grey dashed line shows a power-law relation with a
    given index.
    }
    \label{fig:role-of-halo}
\end{figure*}

The fountain pattern established for the fiducial ensemble provides a reference 
for understanding the baryon cycle quantitatively in the ecosystems of other 
galaxy populations. One expectation is that the overall configuration of the 
fountain pattern, as quantified by the stacked fields of $n$, $T$, $\bm v$, 
$\lambda$, and $I$, is preserved across galaxies with different $M_*$. 
However, the physical extent of the fountain may change with $M_*$, as galaxies 
with higher $M_*$ typically reside in halos with monotonically 
larger $M_{\rm h}$ \cite{yangConstrainingGalaxyFormation2003,yangEvolutionGalaxyDarkMatter2012}, 
leading to larger $R_{\rm h}$.
Given that the stellar size of a galaxy is roughly proportional to 
$R_{\rm h}$ \cite{moFormationGalacticDiscs1998,hearinClusteringConstraintsRelative2019,
marshallDarkagesReionizationGalaxy2019}, and that the ratio between cool-gas 
size and stellar size remains roughly constant across different $M_*$ 
(e.g. Figure~5 of \cite{panRoleRegulatingSize2021}), we expect that the 
scales of the stellar component, the ISM, and the CGM all positively correlate 
with halo size. Thus, we can express this relationship as:
$R_* \propto R_{\rm ISM} \propto R_{\rm CGM} \propto R_{\rm h}$,
at least to the first order.

In Figure~\ref{fig:mdependence}, we present the stacked fields of the six 
gas properties for three ensembles of galaxies at $z=0$ with
$M_* \approx 10^9$, $10^{10}$, and $10^{11} \Msun$, respectively (see 
Table~\ref{tab:subsamples} for the ensembles). To ensure 
fair comparison of fields at different $M_*$, the piecewise-linear scaling 
for the radial coordinate ($r$) in all panels is fixed to that of the fiducial 
ensemble, and the color-coding for each property is unified across all masses.
As anticipated, the stacked fields exhibit configurations closely resembling 
those found for the fiducial ensemble, all displaying distinct features across 
the three radial regimes and three angular zones. The scales of the ISM, CGM 
and IGM increase with rising $M_*$, indicating that more massive galaxies 
acquire gas from larger surrounding regions and respond via feedback to a greater 
extent. Table~\ref{tab:subsamples} lists the $R_{\rm ISM}$, 
$R_{\rm CGM}$ and $R_{\rm h}$ for these ensembles, revealing positive 
correlations among them (Figure~\ref{fig:role-of-halo}G). 
Thus, the stacked maps support a unified fountain 
pattern of the baryon cycle around disk-dominated galaxies over a wide range 
of $M_*$, and the conclusions drawn above for the fiducial ensemble 
qualitatively apply to other ensembles.
In the following, we discuss several points that are particularly useful for 
understanding the mass dependence of the baryon cycle. To supplement this 
discussion, we present the radial profiles of gas properties along the 
disk plane and the outflow cone (Figure~\ref{fig:mdependence-curves}A--F), 
and the angular profiles at three representative radii 
(Figure~\ref{fig:mdependence-curves}G--L).

The quantified mass dependence of gas properties allows us to establish 
scaling relations between baryonic and dark matter components 
semi-empirically. The stellar mass-halo mass 
relation follows approximately $M_{\rm h} \propto M_*^{0.5}$
\cite{zhangMassiveStarformingGalaxies2022}, while the gas mass-stellar mass 
relation is approximately $M_{\rm g} \propto M_*^{0.5}$ \cite{panRoleRegulatingSize2021}. 
In the regime where dark matter dominates gravity so that rotation curves are flat, 
the rotation velocity scales as:
\begin{equation}
    v_\phi \propto V_{\rm h} \propto M_{\rm h}^{1/3}
    \label{eq:v-phi-scaling}
\end{equation}
at fixed redshift. Figure~\ref{fig:mdependence-curves}(B) confirms 
this expectation: the peaks of $v_\phi$ in the disk plane
at $M_* \approx 10^9$, $10^{10}$ and $10^{11}\Msun$ are approximately 
$50$, $100$ and $200\kms$, respectively. Throughout the CGM 
($R_{\rm ISM} \lesssim r \lesssim R_{\rm CGM}$), the $v_\phi$ profile 
follows $r^{-1}$ at all masses, consistent with approximately conserved 
angular momentum (see Appendix~\ref{app:ssec:discussion-am}).

For dynamically cold galactic disks supported by cool gas 
(temperature $T \sim 10^4\,{\rm K}$, sound speed $c_{\rm s} \sim 10\kms$) with 
contributions from transonic turbulence, gas velocity dispersion is 
$\sigma_{\rm g} \sim c_{\rm s}$. A pressure-gravity balance yields a disk scale 
height of:
\begin{equation}
    H_{\rm g} \sim r \frac{\sigma_{\rm g}}{v_\phi} 
    \propto R_{\rm ISM}  \frac{c_{\rm s}}{v_\phi} \frac{r}{R_{\rm ISM}}
    \propto R_{\rm h} \frac{1}{ V_{\rm h} } \frac{r}{R_{\rm ISM}}
    \propto \frac{r}{R_{\rm ISM}}\,.
    \label{eq:hg-r}
\end{equation}
This scaling demonstrates that disk height, 
at a fixed fraction of $R_{\rm ISM}$, is independent of $M_*$ at fixed redshift. 
Figure~\ref{fig:mdependence} supports this prediction: the cool-gas 
disks ($T \sim 10^4\,{\rm K}$) appear similarly thin across the full $2\dex$ range 
in stellar mass.

In the CGM, disk-plane temperature profiles remain approximately isothermal, 
increasing from $10^{4.8}$--$10^{5}{\rm K}$ at $M_* = 10^9\Msun$ to 
$10^{5.5}$--$10^{6}{\rm K}$ at $M_* = 10^{11}\Msun$. This $\approx 1\dex$ 
increase matches the expected virial-temperature scaling 
$T_{\rm h} \propto M_{\rm h}^{2/3} \propto R_{\rm h}^2$. 
At $r \approx 1\Mpc$, a sharp temperature drop separates the IGM 
environment from the CGM environment of the ensemble with 
$M_* \lesssim 10^{10}\Msun$: warm-hot gas ($T \gtrsim 10^5\,{\rm K}$) heated by 
virialization shocks surrounds galaxies, while cool gas ($T \lesssim 10^4\,{\rm K}$) 
fills the remaining, more void volume 
($n \lesssim 10^{-7} \perccm \approx \rho_{\rm c} \Omega_{\rm B,0}$; 
see \cite{liELUCIDVIIUsing2022}). 
More massive galaxies show extended warm-hot gas to larger scales, 
reflecting their residence in larger structures. This separation scale coincides 
with the transition between one-halo and two-halo regimes found in clustering 
analyses \cite{shiMappingRealspaceDistributions2016}, providing a 
complementary probe of large-scale structure.

The average gas density in the disk, within a fixed fraction of $R_{\rm ISM}$,
can be estimated as:
\begin{equation}
    \rho_{\rm g} 
    \propto \frac{M_{\rm g}}{ H_{\rm g} R_{\rm ISM}^2 }
    \propto \frac{M_*^{0.5}}{R_{\rm h}^2} 
    \propto \frac{M_{\rm h}}{R_{\rm h}^2} 
    \propto M_{\rm h}^{1/3} 
    \,.\label{eq:rho-mstar}
\end{equation}
This positive mass dependence explains the observed increase in density 
with $M_*$ (Figure~\ref{fig:mdependence-curves}A). For IllustrisTNG 
galaxies, this scaling also implies a dependence of gas fraction on SMBH mass:
\begin{equation}
    f_{\rm g} \equiv \frac{M_{\rm g}}{M_*}
    \propto \frac{M_*^{0.5}}{M_*} 
    \propto M_*^{-0.5}
    \propto M_{\rm BH}^{-0.5}\,,
\end{equation}
using the relation $M_{\rm BH} \propto M_*$ that was found for IllustrisTNG \cite{liPhysicalProcessesCoevolution2025,habouzitSupermassiveBlackHoles2021}.
This negative correlation with exponent $-0.5$ is indeed consistent with recent 
observations \cite{wangBlackHolesRegulate2024}.

The disk-plane spin profile exhibits a remarkably universal form across all 
stellar masses, indicating a self-similarity of the rotational 
motion in the baryon cycle.
The profile has a rising segment in the inner ISM, 
a peak near the ISM-CGM interface, and a flat segment in the outer CGM,
which can thus be described by
eq.~\ref{eq:model-lambda-disk} with $\tilde{r}_\lambda$, $\lambda_{\rm p}$
and $\lambda_0$ being nearly mass-independent.
The universal spin profile has been found for dark matter 
in N-body simulations \cite{bullockUniversalAngularMomentum2001},
which, in combination with a conservation of angular momentum during 
gas transport in the CGM, and a universal angular-momentum
redistribution in the ISM, may explain the universality of the 
gas spin profile.
In the IGM, large fluctuations of the spin profile indicate a
decoupling of gas motion in galaxy ecosystems from the environment 
at large scales.

Disk-plane radial velocity and mass flux profiles (Figure~\ref{fig:mdependence-curves}E,F) reveal that more massive galaxies 
maintain stronger, faster and more spatially extended inflows within the 
CGM regime -- a requirement for sustaining 
higher star formation rates. The shapes of the inflow profiles 
are similar across masses, indicating that the inflows are driven by the 
same set of mechanisms, and motivating the use of 
a universal function to parameterize the profile (eq.~\ref{eq:model-v-r-disk}).
At $M_* = 10^{11}\Msun$, inflow reaches peak 
values of $-v_r \sim 30\kms$ and $-I \sim 0.5\Msun{\rm yr}^{-1}{\rm sr}^{-1}$ 
at $r \approx 100\Kpc$. The inflow decelerates at smaller radii and reverses 
near $R_{\rm ISM}$ due to angular-momentum barriers, confirming that infalling 
gas accumulates at the CGM-ISM interface while awaiting 
angular-momentum redistribution.

The anisotropy between disk-plane and outflow-cone gas properties 
(Figure~\ref{fig:mdependence-curves}) reflects the 
interplay between disk flattening and feedback processes. 
Outflow-cone density profiles (panel A) remain similar across stellar 
masses, suggesting a higher fraction of gas depletion from more massive 
halos. Temperature differences between outflow cone and disk plane at 
$r < R_{\rm ISM}$ (panels D, J) increase with $M_*$ due to enhanced 
virialization heating in more massive systems. Temperature anisotropy remains negligible at CGM scales across all masses, reinforcing the 
conclusion that feedback-driven outflows produce only subdominant heating
in the CGM of our IllustrisTNG ensembles.
The mass flux becomes nearly isotropic at $r \approx 10$--$30\Kpc$ 
around the ensemble with $M_* = 10^{11}\Msun$ (panel F),
implying a dynamically hot gas environment. The isotropic outflow
also means that the ISM undergoes a net mass loss, and that
the galaxies are on a path toward quenching, unless fresh gas is
supplied in a clumpier form (e.g. through gas clumps or mergers) 
that was filtered out by the median operation in the stacking.
The density drop in the inner ISM at $M_* = 10^{11}\Msun$ 
(panel A) verifies this ongoing quenching process, and is consistent with 
the inside-out quenching scenario \cite{liPMaNGAGradientsRecent2015,linSDSSIVMaNGAInsideout2019,nelsonSpatiallyResolvedStar2021}.

Figure~\ref{fig:role-of-halo} summarizes the dependence of ecosystem divisions 
($R_{\rm ISM}$ and $R_{\rm CGM}$) and gas properties at representative
radii in ISM and CGM on halo mass, together with the scaling relations derived 
above. These trends underscore the fundamental role of dark matter halos in shaping 
baryon cycles, and support the use of halo properties to scale the 
baryonic properties, thereby simplifying the parametric modeling presented 
in Appendix~\ref{app:ssec:model}.


\subsection{\label{app:ssec:redshift-dependence}Redshift dependence of baryon cycle}

\begin{figure*}[t]
\centering
\includegraphics[width=\textwidth]{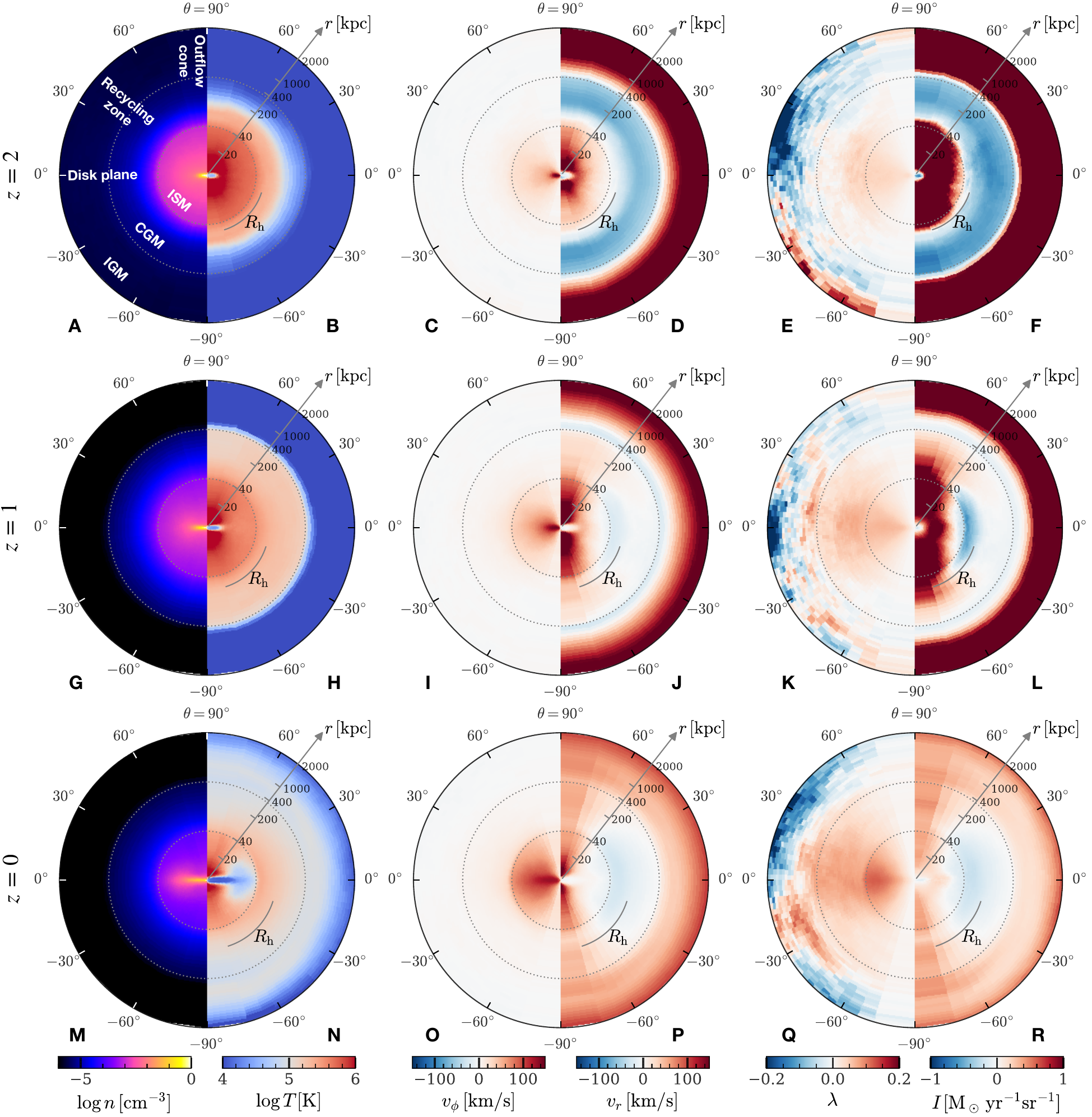}
\caption{
    {\figem Stacked 2D maps of gas properties around ensembles of galaxies 
    at different redshifts.}
    This figure is similar to the first row of Figure~\ref{fig:mass10z0}, 
    but comparing gas properties across three ensembles of star-forming 
    galaxies with $M_* \approx 10^{10} \Msun$ at 
    {\figem (A)}--{\figem (F)}, $z = 2$, 
    {\figem (G)}--{\figem (L)}, $z = 1$, and 
    {\figem (M)}--{\figem (R)}, $z = 0$. 
    The piecewise-linear radial scaling and annotated ISM, CGM and IGM regimes 
    are fixed to match the $z=0$ ensemble.
    The pattern of baryon cycle persists across cosmic time, though physical 
    extents expand with decreasing $z$.
    At higher $z$, galaxy ecosystems exhibit dynamically hotter configurations 
    with stronger, more isotropic inflow and outflow.
    See Figure~\ref{fig:zdependence-curves} for the corresponding 
    1D profiles.
    }
    \label{fig:zdependence}
\end{figure*}

\begin{figure*}[t] 
	\centering
    \includegraphics[width=\textwidth]{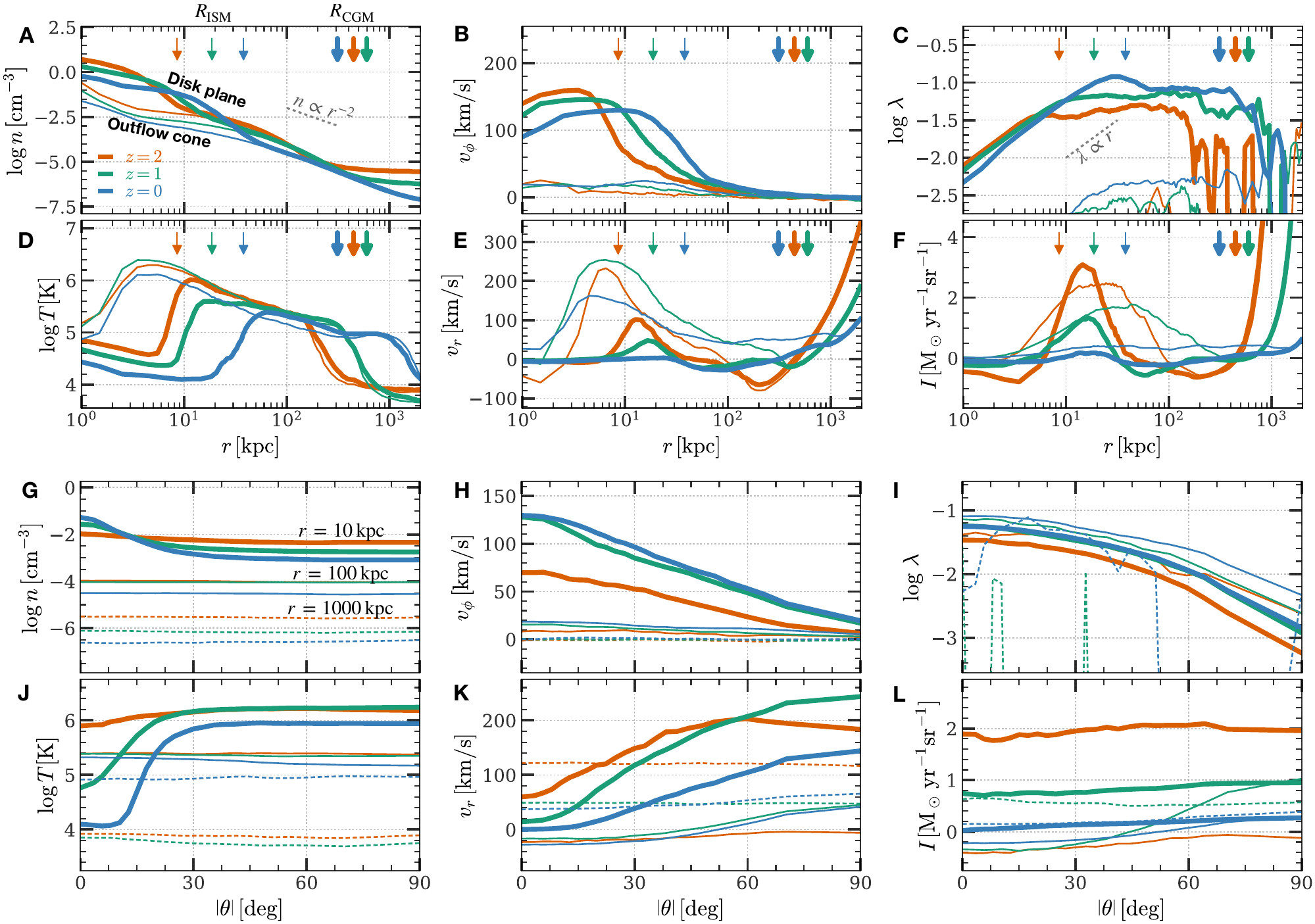}
    \caption{
    {\figem Stacked 1D profiles of gas properties around ensembles of 
    galaxies at different redshifts.}
    {\figem (A)}--{\figem (F)} Radial profiles along the disk plane 
    (thick curves) and outflow-cone center (thin curves).
    {\figem (G)}--{\figem (L)} Angular profiles at three representative radii.
    Thin and thick arrows mark the ISM-CGM interface ($R_{\rm ISM}$) and CGM-IGM 
    interface ($R_{\rm CGM}$), respectively, for each ensemble.
    See Figure~\ref{fig:zdependence} for the 
    corresponding 2D maps. 
    }
    \label{fig:zdependence-curves}
\end{figure*}

The fountain pattern found in the ensembles at $z \approx 0$ should persist at 
higher $z$, but on different physical scales, given the role of halos 
in shaping the baryon cycle. The halo radius depends 
on $M_{\rm h}$ and $z$ as:
\begin{equation}
    R_{\rm h} = \left[ \frac{2GM_{\rm h}}{\Delta_{\rm v}} \right]^{1/3} H(z)^{-2/3}
    \,.
    \label{eq:rh-h}
\end{equation}
Since the stellar mass-halo mass relation evolves only mildly with 
redshift \cite{mosterEMERGEEmpiricalModel2018,behrooziUNIVERSEMACHINECorrelationGalaxy2019}, 
ecosystems at fixed $M_*$ should have spatial scales that increase by 
a factor of a few from $z = 2$ to $z = 0$.
We test this prediction using stacked fields at three redshifts ($z = 0$, $1$, $2$) 
for galaxies at a fixed $M_* \approx 10^{10}\Msun$ (Table~\ref{tab:subsamples}, IDs=2, 4, 5). 
Figures~\ref{fig:zdependence} and \ref{fig:zdependence-curves} show that the 
fountain pattern is preserved at all $z$, with clear ISM, CGM and IGM 
divisions evident in both the 2D maps and 1D profiles.

From $z = 2$ to $z = 0$, $R_{\rm ISM}$ increases from $\approx 10\Kpc$ to 
$\approx 40\Kpc$, tracking the growth of $R_{\rm h}$. 
The CGM radius $R_{\rm CGM}$ shows non-monotonic evolution due to the competition 
between gravity and cosmic expansion. Gas density in the inner ISM 
decreases with decreasing redshift, reflecting the expansion 
of galaxy sizes.
Inner CGM temperatures decline by $\sim 0.5\dex$ from $z = 2$ to $z = 0$, 
consistent with the redshift dependence of virial temperature,
$T_{\rm h} \propto V_{\rm h}^2 \propto M_{\rm h}^{2/3} H(z)^{2/3}$. 
Disk temperatures also decrease substantially due to reduced density and the 
eEOS applied to star-forming gas in IllustrisTNG.
The maximum rotation velocity decreases from $\sim 160\kms$ at $z = 2$ to 
$\sim 130\kms$ at $z = 0$. 
Conversely, the CGM spin $\lambda$ increases at lower $z$ 
(Figure~\ref{fig:zdependence-curves}C), 
indicating that gas acquires more angular momentum during its infall in the 
low-$z$ Universe. 
This trend aligns with predictions that angular momentum accumulation 
is a time-integrated consequence of tidal torques \cite{stewartAngularMomentumAcquisition2013,
danovichFourPhasesAngularmomentum2015}.
Despite these redshift dependencies, the spin profile at all epochs follows 
the universal form found at $z = 0$.

High-redshift galaxies exhibit notably different gas dynamics.
The ISM appears thicker and the CGM is more isotropic at $z = 2$ than at $z = 0$ 
(visible in the 2D maps in Figure~\ref{fig:zdependence}). 
This ``hotter'' dynamical state arises from two factors: 
(i) lower spin in the CGM due to insufficient time 
for tidal torque to accumulate angular momentum, 
as seen from the stacked profiles (Figure~\ref{fig:zdependence-curves}C);
(ii) higher halo accretion rates that drive violent dynamical heating,
as suggested by \cite{moTwophaseModelGalaxy2024} and \cite{mccluskeyDiscSettlingDynamical2024},
with additional contributions from the filamentary geometry of accretion \cite{dekelColdStreamsEarly2009} and 
bursty star formation \cite{el-badryBreathingFIREHow2016}, both injecting turbulent energy into the gas.
The hotter dynamics at high $z$ fundamentally alters outflow morphology. 
At $z = 2$, outflows are more isotropic at the ISM-CGM interface, 
with substantial mass flux along the disk plane (comparable to or exceeding 
the polar outflow flux; see Figure~\ref{fig:zdependence-curves}F
at $r \approx 10$--$20\Kpc$). 
This contrasts with the bipolar geometry at $z = 0$, which, as discussed above, 
emerges from the interaction of feedback-driven outflows with an anisotropic 
ambient medium. 
Thus, for star formation to be sustained, bulk inflow around galaxies at $z=2$ 
must be in clumpy form (e.g. through cold-gas filaments or mergers) that was 
filtered out by the median operation in the stacking.

\subsection{\label{app:ssec:discussion-bh}Implications for the growth of supermassive black holes}

\begin{figure*}[t] 
	\centering
    \includegraphics[width=\textwidth]{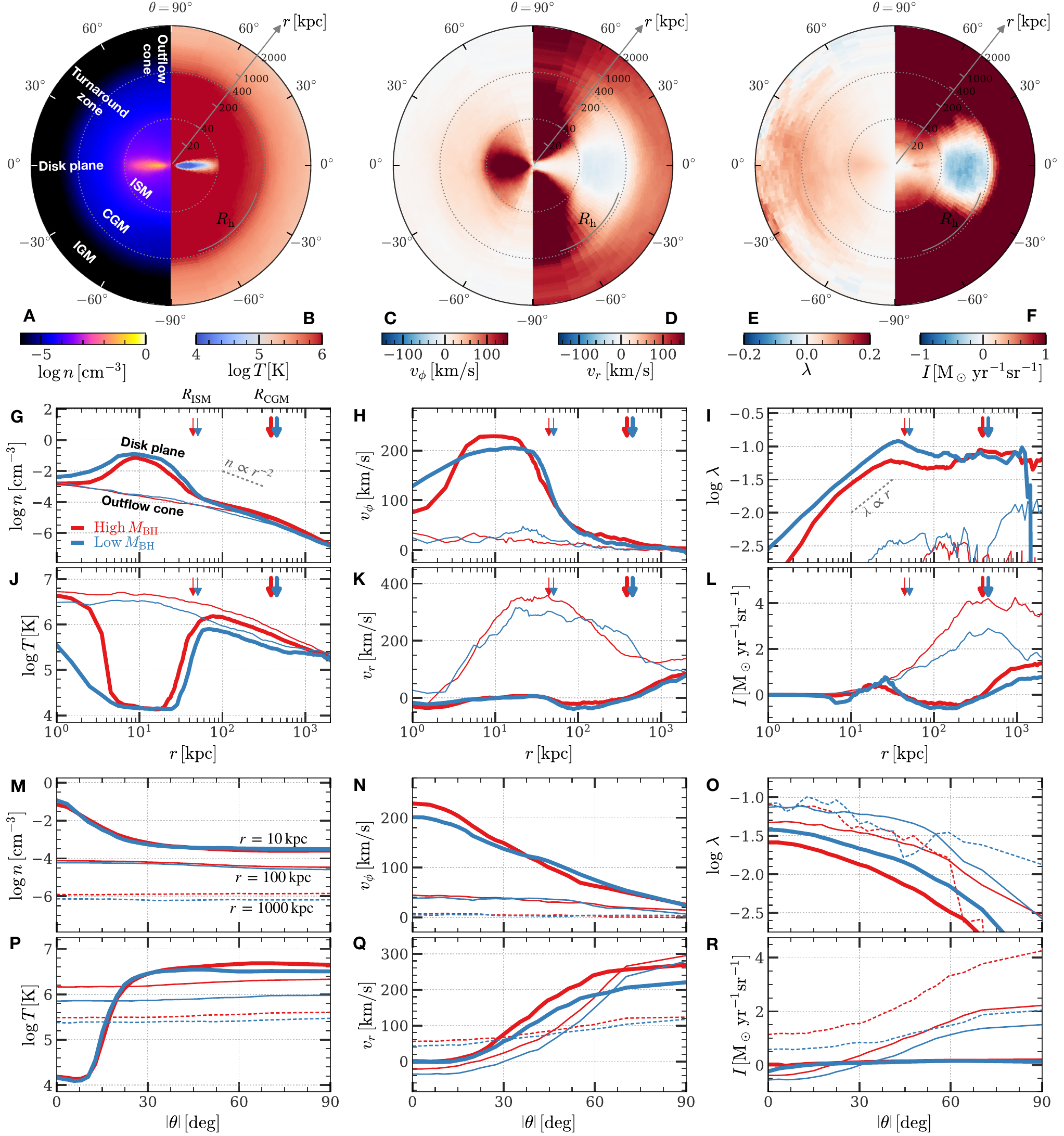}
    \caption{
    {\figem Stacked 2D maps and 1D profiles of gas properties around an ensemble
    of galaxies hosting massive SMBHs.}
    {\figem (A)}--{\figem (F)} 2D maps of gas properties.
    {\figem (G)}--{\figem (L)} Radial profiles along the disk plane 
    (thick curves) and outflow-cone center (thin curves).
    {\figem (M)}--{\figem (R)} Angular profiles at three representative radii.
    The ensemble is constructed using the same selection criteria as ensemble 3 
    (star-forming galaxies at $z=0$ with $M_*\approx 10^{11}\Msun$), 
    except with the requirement $M_{\rm BH} > 10^{8.2}\Msun$. 
    For comparison, blue curves in the 1D-profile panels show the ensemble with $M_{\rm BH} \leqslant 10^{8.2}\Msun$ 
    (ensemble 3). 
    }
    \label{fig:highmassSMBH}
\end{figure*}

One application of the stacking method is to investigate the baryon cycles 
around galaxies with massive SMBHs. These systems are expected to arise in
dynamically hot CGM and ISM environments in which 
enhanced SMBH accretion and AGN feedback can occur 
\cite{moTwophaseModelGalaxy2024,chenTwophaseModelGalaxy2025a},
and their baryon cycles are thus expected to differ from those of galaxies with 
less massive SMBHs.
In Figure~\ref{fig:highmassSMBH}, we present stacked fields of gas properties 
around galaxies hosting massive SMBHs ($M_{\rm BH} > 10^{8.2} \Msun$), 
with other selection criteria identical to ensemble 3 
(star-forming, $z = 0$, $M_*\approx 10^{11}\Msun$; see Table~\ref{tab:subsamples}).
Radial and angular profiles (Figure~\ref{fig:highmassSMBH}G--L, M--R, red curves) 
reveal pronounced differences compared to low-$M_{\rm BH}$ systems (blue curves).
The baryon distribution around high-$M_{\rm BH}$ galaxies is more compact, 
evidenced by smaller $R_{\rm ISM}$ and $R_{\rm CGM}$ values (marked by arrows). 
The density profile still shows a rapid rise at the ISM-CGM interface, 
indicating that the angular-momentum barrier remains in place to decelerate 
infalling gas and flatten gas geometry. 
However, at $r \lesssim 10\Kpc$, the high-$M_{\rm BH}$ ensemble shows a larger 
density drop and a much larger temperature increase than 
the low-$M_{\rm BH}$ ensemble, indicating stronger AGN feedback.
The $v_\phi$ profile at $r\approx 10\Kpc$ is significantly elevated compared to 
the low-$M_{\rm BH}$ ensemble, indicating an elevated dark-matter density
in the inner halos.
This is consistent with the larger halo radius ($R_{\rm h} = 291.4\Kpc$, compared to $234.7\Kpc$ for low-$M_{\rm BH}$ systems) 
and correspondingly larger halo mass. Notably, the spin profile is substantially 
reduced, indicating that the CGM supplies slowly-spinning gas to these galaxies. 
Combined with the fact that a more massive halo provides a denser environment
with more frequent close encounters/mergers, the enhanced SMBH activity
may be explained \cite{rodriguez-gomezRoleMergersHalo2017,
duPhysicalOriginMasssize2024,maEvolutionaryPathwaysDisk2024}. 
We note that the interaction between SMBH activity and gas spin is 
bidirectional: the ejective effect of AGN feedback may preferentially remove 
low-spin gas \cite{zjupaAngularMomentumProperties2017} -- a response 
to SMBH growth that warrants further investigation.
AGN feedback also enhances outflow speed and flux 
throughout the CGM (panels L,R), indicating an ejective nature of 
AGN feedback at high $M_{\rm BH}$. 

These findings are also related to the redshift dependence of baryon cycles 
presented above in Appendix~\ref{app:ssec:redshift-dependence}.
Since angular momentum plays a fundamental role in limiting gas accretion onto SMBHs 
in the inner ISM \cite{hobbsFeedingSupermassiveBlack2011,
gaspariChaoticColdAccretion2015,moTwophaseModelGalaxy2024,
chenTwophaseModelGalaxy2024,chenTwophaseModelGalaxy2025a},
and turbulence can efficiently generate low-angular-momentum gas elements 
that fuel SMBH accretion (analogous to the role of the ISM-CGM interface in 
sifting slowly-rotating gas), the dynamically hotter environments 
of high-$z$ galaxies are expected to promote SMBH accretion. This trend may partially 
explain elevated AGN activity at high $z$
\cite{weinbergerSupermassiveBlackHoles2018}.

We acknowledge important limitations in the current sub-grid model implemented in IllustrisTNG. 
SMBH accretion follows the Bondi-Hoyle prescription \cite{bondiMechanismAccretionStars1944,
bondiSphericallySymmetricalAccretion1952}, 
which is certainly too simplified to account for effects such as the barrier of 
angular momentum. The eEOS artificially thermalizes and isotropizes gas motion at 
small scales, obscuring the turbulent cascades that naturally 
boost SMBH growth \cite{vollmerQuenchingStarFormation2013}. Additionally, feedback from both 
stars and AGNs is injected isotropically at small scales, relying on subsequent 
hydrodynamic coupling to generate anisotropy \cite{pieriAnisotropicGalacticOutflows2007,
pinsonneaultAnisotropicGalacticOutflows2010}. While these simplifications are necessary for 
simulating galaxies over large volumes, zoom-in simulations with extensive coverage in the 
parameter space and carefully calibrated sub-grid models offer a promising path 
forward for overcoming these limitations.

\subsection{\label{app:ssec:formulation}A generalized formulation of the stacking method}

The stacking method was used to reveal the regular patterns in individual 
gas properties for galaxy ensembles. This approach has two key 
limitations. First, it captures only the median 
of a property at each local-frame position, overlooking the 
multiphase structures that may coexist there. Examples of such structures
include unmixed inflow and outflow components, cold and hot gas phases 
within outflows, and gas with different chemical compositions. 
Such multiphase structures have been reported in phase-space distributions of simulated galaxies 
(e.g. Figures~9,10 of \cite{nelsonFirstResultsTNG502019}). 
Second, the method stacks individual properties separately and thus cannot 
capture correlations between the properties at a given location,
thereby missing the opportunity to uncover the underlying mechanisms that shape 
the correlations.

The stacking method can be generalized to overcome these limitations if 
we can establish the joint distribution $p(\bm x, \bm q)$ of local-frame 
coordinate $\bm x$ and the set of baryonic properties $\bm q$ in question.
The conditional distribution 
$p(\bm q | \bm x)$ then contains the full information needed to identify the 
multiphase structures of gas coexisting at $\bm x$, while the marginal distributions 
of pairs of properties encode their correlations. 
This generalized stacking formulation can then be expressed as:
\begin{equation}
    f = \mathcal{S}\left[ p(\bm x, \bm q) \right] \,,
\end{equation}
where the stacking functional $\mathcal{S}$ transforms the joint distribution 
to produce the desired output $f$. For a single property such as temperature 
(${\bm q} = T$), applying an integration operator $\mathcal{S}\left[p\right] = \int p({\bm x}, T) T dT / p({\bm x})$ 
recovers the mean temperature at $\bm x$, reducing to the standard 
stacking method.
Our original formulation of stacking 
represents a special case of this general approach: it samples $p(\bm x, \bm q)$ to obtain gas elements 
around individual galaxies in an ensemble, performs averaging to form super-cells 
around each galaxy, and takes the median among these super-cells 
to obtain the composite field.
With an ensemble sufficiently large to establish the full distribution 
$p(\bm x, \bm q)$, or at least its approximation via numerical sampling, 
clustering analysis can be applied to identify 
phase-space structures. This technique was used to decompose stars in 
individual galaxies into distinct kinematic components such as disk and bulge
\cite{duKinematicDecompositionIllustrisTNG2020,zanaMorphologicalDecompositionTNG502022,
liangConnectionGalaxyMorphology2025,maTwophaseFormationGalaxies2026},
and our generalized formulation provides a natural extension in which
an analogous decomposition can be made for cycling gas in galaxy ecosystems.

\subsection{\label{app:ssec:metal}Spatial pattern of chemical enrichment}

\begin{figure*}[t] 
	\centering
    \includegraphics[width=0.825\textwidth]{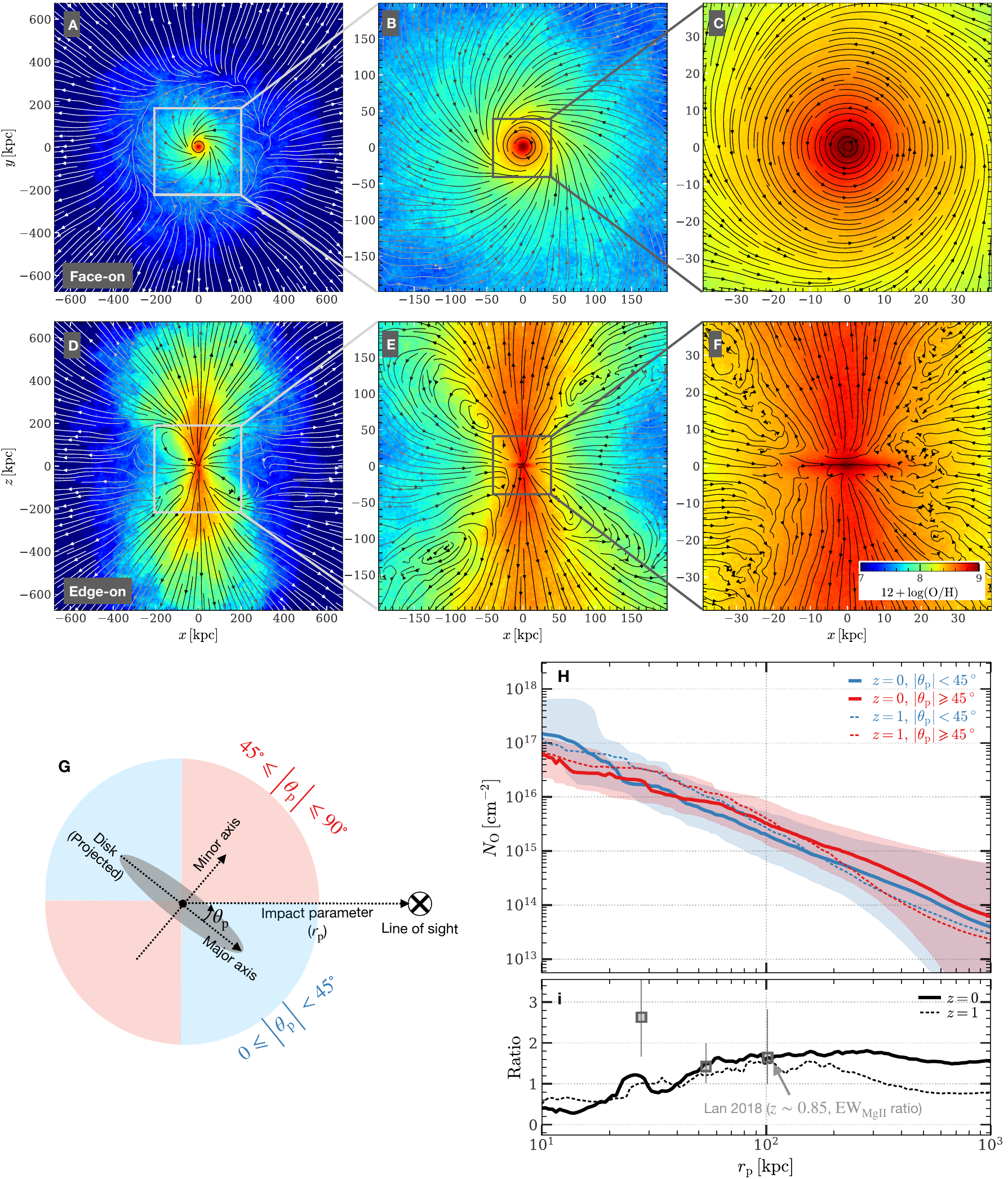}
    \caption{
    {\figem Metallicity field around ensembles of galaxies.}
    {\figem (A)}--{\figem (F)} Stacked metallicity field 
    around the fiducial ensemble ($z=0$),
    viewed face-on ({\figem A}--{\figem C}) and edge-on 
    ({\figem D}--{\figem F}) at three zoom-in levels.
    The slices shown here are the same as those in 
    Figure~\ref{fig:streamline}, with the velocity field 
    overplotted as streamlines for reference.
    {\figem (G)} Projection scheme used to evaluate the metal column density 
    along randomly sampled lines of sight.
    {\figem (H)} Oxygen column density ($N_{\rm O}$) as a function of 
    projected distance
    ($r_{\rm p}$) for different ranges of projected inclination angle 
    ($\theta_{\rm p}$) around the fiducial ensemble (solid
    curves with shading). 
    Results for ensemble 4 ($z=1$) are shown for comparison 
    (dashed curves). 
    {\figem (I)} Ratio between $N_{\rm O}$ evaluated at 
    $\left|\theta_{\rm p}\right| \geqslant 45^\circ$ 
    and $\left|\theta_{\rm p}\right| < 45^\circ$.
    For comparison, we also show the ratio of the equivalent width (EW)
    of Mg{\sc II} absorption between the same two $\theta_{\rm p}$ ranges,
    as obtained by \cite{lanCircumgalacticMediumEBOSS2018}
    for emission-line galaxies with comparable $M_*$
    at $z \sim 0.85$.
    In {\figem (H)} and {\figem (I)}, curves and markers show the median,
    and shading and error bars indicate the $16^{\rm th}$--$84^{\rm th}$ 
    percentile range.
    }
    \label{fig:metal-field}
\end{figure*}

We obtained the composite metallicity field using the same stacking 
procedure as for other gas properties.
For our fiducial ensemble, we show the stacked field of metallicity 
(expressed as 12 plus logarithmic ratio of oxygen-to-hydrogen number) 
in Figure~\ref{fig:metal-field}(A)--(F) 
in the same slices as Figure~\ref{fig:streamline}, and we overplot the 
streamlines of the stacked velocity field for reference.
The metallicity field exhibits a spatial pattern that closely follows
the fountain structure revealed by the velocity field. 
Specifically, an enriched disk with solar-level metallicity is visible
within the ISM regime, which serves as the source of metals for the 
entire ecosystem; a cone-shaped zone of elevated metallicity 
is seen within the CGM regime, which aligns with the bipolar outflow 
and signifies the role of feedback-driven outflow in 
the transport of metals;
the zone around the disk plane within the CGM regime
has lower metallicity than the outflow cone, reflecting the weaker
outflow-driven metal transport and stronger inflow-driven dilution.
The similarity between the metallicity and velocity fields implies
that the fountain pattern can be revealed by both kinematic and chemical 
tracers, and indicates that observations can combine different tracers 
to place stringent constraints on the baryon cycle 
\cite{chenCosmicEvolutionSpatial2025,chenCircumgalacticMediumTraced2025} 
(see also Lyu et al. 2026 in prep).

To test whether the anisotropy of the metal distribution is observationally 
detectable, we randomly sampled sightlines, each through the ecosystem around 
a randomly selected galaxy in a given ensemble. We then integrated the 
density field of a given species $\rm X$ along each sightline, with an 
integration depth of $\pm1500\Kpc$,
to obtain the column density (number of metal atoms per unit area) of that species, 
$N_{\rm X}$. The plane perpendicular to the sightline is parameterized by 
$r_{\rm p}$, the projected distance (impact parameter) from the sightline to 
the galaxy center, and $\theta_{\rm p}$, the projected inclination angle
of the sightline with respect to the major axis of the projected disk.
The variation of $N_{\rm X}$ with $r_{\rm p}$ and $\theta_{\rm p}$ 
thus encodes the inhomogeneity and anisotropy of the metal distribution, 
with projection effects taken into account 
(see Figure~\ref{fig:metal-field}G for a schematic illustration).
In Figure~\ref{fig:metal-field}(H), we show the oxygen column density, $N_{\rm O}$ 
(median and percentiles among sightlines), as a function of $r_{\rm p}$ in two 
ranges of $\theta_{\rm p}$, $\left|\theta_{\rm p}\right| < 45^\circ$
and $\left|\theta_{\rm p}\right| \geqslant 45^\circ$, for the fiducial ensemble.
In Figure~\ref{fig:metal-field}(I), we show the ratio between the median 
$N_{\rm O}$ in the two $\theta_{\rm p}$ ranges. 
At $r_{\rm p} \gtrsim 40\Kpc$, the ratio is above unity, 
consistent with the elevated metallicity in the outflow cone.
The ratio is below unity at $r_{\rm p} \lesssim 20\Kpc$, consistent with the 
high metallicity around the disk plane.
For comparison, we also show the results for ensemble 4 ($z=1$; 
see Table~\ref{tab:subsamples}),
which exhibits similar anisotropy.

In observations, the anisotropy of metal distribution can be inferred from 
absorption-line spectroscopy of background sources, as demonstrated 
by ref.~\cite{lanCircumgalacticMediumEBOSS2018} using background quasars 
for a sample of emission-line galaxies at $z \sim 0.85$.
Measurements of Mg{\sc II} equivalent width (EW) at different $\theta_{\rm p}$ 
show an anisotropy pattern similar to our prediction, 
as indicated by the markers in Figure~\ref{fig:metal-field}(I).
However, the conversion from EW to column density is non-trivial, especially
given the high column densities and complex phase-space structures of 
baryons expected in the inner ecosystems.
A forward-modeling approach is therefore needed to predict EW while taking 
these factors into account \cite{nelsonSyntheticAbsorptionLine2025}.

\end{appendix}

\end{multicols}
\end{document}